\documentclass[iop, apj, natbib]{emulateapj}
\usepackage{graphicx}
\usepackage{amsmath}
\usepackage{amssymb}
\usepackage{color}
\usepackage{multirow}
\usepackage{rotating}
\usepackage{booktabs}
\usepackage{times}
\usepackage{natbib}

\usepackage{comment}

\newcommand{\zi}{\textit{z}}

\newcommand{\lya}{Ly$\alpha$}

\newcommand{\xhi}{$x_{\textsc{Hi}}$}
\newcommand{\tel}{$\tau_\mathrm{el}$}

\slugcomment{Submission : April 24, 2014, Accepted : September 16, 2014}

\shorttitle{Accelerated Evolution of the \lya\ Luminosity Function at $z\gtrsim 7$}
\shortauthors{Konno et al.}

\begin{document}

\title{Accelerated Evolution of the \lya\ Luminosity Function at $\zi \gtrsim 7$ Revealed by \\
 the Subaru Ultra-Deep Survey for \lya\ Emitters at $\zi=7.3$}

\author{Akira Konno\altaffilmark{1,2}, 
	Masami Ouchi\altaffilmark{1,3},
	Yoshiaki Ono\altaffilmark{1}, 
	Kazuhiro Shimasaku\altaffilmark{2,4}, 
	Takatoshi Shibuya \altaffilmark{1,5},
	Hisanori Furusawa \altaffilmark{6}	\\
	Kimihiko Nakajima\altaffilmark{2},
	Yoshiaki Naito \altaffilmark{1},
	Rieko Momose\altaffilmark{1},
	Suraphong Yuma \altaffilmark{1} and
	Masanori Iye \altaffilmark{6}
        }

\altaffiltext{1}{Institute for Cosmic Ray Research, The University of Tokyo, Kashiwa-no-ha, Kashiwa 277-8582, Japan}
\altaffiltext{2}{Department of Astronomy, Graduate School of Science, The University of Tokyo, Hongo, Bunkyo-ku, Tokyo, 113-0033, Japan}
\altaffiltext{3}{Kavli Institute for the Physics andMathematics of the Universe (Kavli IPMU), WPI, The University of Tokyo, Kashiwa, Chiba 277-8583, Japan}
\altaffiltext{4}{Research Center for the Early Universe, Graduate School of Science, The University of Tokyo, Hongo, Bunkyo-ku, Tokyo, 113-0033, Japan}
\altaffiltext{5}{Center for Computational Science, University of Tsukuba, Tsukuba, Ibaraki 305-8577, Japan}
\altaffiltext{6}{National Astronomical Observatory of Japan, 2-21-1 Osawa, Mitaka, Tokyo 181-8588, Japan}

\email{konno@icrr.u-tokyo.ac.jp}

\begin{abstract}

We present the ultra-deep Subaru narrowband imaging survey for \lya\ emitters (LAEs) at $\zi=7.3$
in SXDS and COSMOS fields ($\sim 0.5$ deg$^2$)
with a total integration time of 106 hours. 
Exploiting our new sharp bandwidth filter, $\textit{NB101}$,
installed on Suprime-Cam, we have reached 
$L( \mathrm{Ly}\alpha ) = 2.4 \times 10^{42} \ \mathrm{erg} \ \mathrm{s}^{-1}$ ($5\sigma$)
for $\zi=7.3$ LAEs, about 4 times deeper than previous Subaru $\zi \gtrsim 7$ studies,
which allows us to reliably investigate the evolution of the \lya\ luminosity function (LF),
for the first time, down to the luminosity limit same as those of Subaru $z=3.1-6.6$ LAE samples.
Surprisingly, we only find three and four LAEs in SXDS and COSMOS fields, respectively,
while one expects a total of $\sim 65$ LAEs by our survey in the case of no \lya\ LF evolution from $\zi = 6.6$ to $7.3$.
We identify a decrease of the \lya\ LF from $\zi =6.6$ to $7.3$ at the $> 90\%$ confidence level
from our $\zi = 7.3$ \lya\ LF with the best-fit Schechter parameters of
$L^{*}_{\mathrm{Ly}\alpha} = 2.7^{+8.0}_{-1.2} \times 10^{42} \ \mathrm{erg} \ \mathrm{s}^{-1}$ 
and $\phi^{*} = 3.7^{+17.6}_{-3.3} \times 10^{-4} \ \mathrm{Mpc}^{-3}$ for a fixed $\alpha = -1.5$.
Moreover, the evolution of the \lya\ LF is clearly accelerated at $\zi > 6.6$ beyond the measurement
uncertainties including cosmic variance.
Because no such accelerated evolution of the UV-continuum LF or the cosmic star-formation rate (SFR)
is found at $z\sim 7$, but suggested only at $z>8$ \citep{2013ApJ...773...75O, 2014arXiv1403.4295B}, 
this accelerated \lya\ LF evolution is explained 
by physical mechanisms different from a pure SFR decrease
but related to the \lya\ production and escape in the process of cosmic reionization. 
Because a simple accelerating increase of IGM neutral hydrogen absorbing \lya\ 
would not reconcile with Thomson scattering optical depth measurements from \textit{WMAP} and \textit{Planck},
our findings may support new physical pictures suggested by recent theoretical studies, 
such as the existence of {\sc Hi} clumpy clouds within cosmic ionized bubbles 
selectively absorbing \lya\ and the large ionizing photon escape fraction of galaxies making weak \lya\ emission.

\end{abstract}



\section{Introduction} \label{intro}

\lya\ emitters (LAEs) are young star-forming galaxies,
and are essential to explore very high redshift universe.
A large number of systematic narrowband imaging surveys
have been carried out for LAEs at $\zi \sim 7$
\citep{2006Natur.443..186I, 2008ApJ...677...12O, 2010ApJ...722..803O, 2012ApJ...752..114S}
and beyond $\zi \sim 8$
\citep{2005MNRAS.357.1348W, 2007A&A...461..911C, 2008MNRAS.384.1039W, 2009MNRAS.398L..68S,
2010A&A...515A..97H, 2010ApJ...721.1853T, 2012A&A...538A..66C, 2012ApJ...745..122K, 2014MNRAS.440.2375M}.
In these studies, it is under debate whether the \lya\ luminosity function (LF) of LAEs evolve from $\zi = 6.6$ or not,
while no evolution of the \lya\ LF in $\zi = 3.1 - 5.7$ \citep{2008ApJS..176..301O} 
and a decrease from $\zi = 5.7$ to $6.6$ \citep{2006ApJ...648....7K, 2011ApJ...734..119K, 2010ApJ...723..869O, 2010ApJ...725..394H}
have been identified.
\cite{2010A&A...515A..97H}, \cite{2010ApJ...721.1853T} and \cite{2012ApJ...745..122K} conclude that
there is no evolution of the \lya\ LF from $\zi = 6.6$ to $7.7$.
On the other hand, \cite{2012A&A...538A..66C} place the upper limit on the \lya\ LF 
based on their result of no detection of $\zi = 7.7$ LAE,
and rule out no evolution of the \lya\ LF in $\zi = 6.6 - 7.7$.
Moreover, the observations for $\zi = 7.0$ and $7.3$ LAEs have been conducted by \cite{2010ApJ...722..803O} and \cite{2012ApJ...752..114S},
and these studies find that the number density and the \lya\ luminosity density decrease from $\zi = 5.7$ to $7.0 - 7.3$.
However, they cannot clearly find whether the \lya\ LF evolves from $\zi = 6.6$ to $7.0 - 7.3$ due to the large uncertainties of 
their LF measurements. Their large uncertainties are originated from the 
relatively shallow imaging that just reaches the bright \lya\ luminosity limit of $L(\mathrm{Ly}\alpha) \sim 10^{43}$ erg s$^{-1}$.
The contradicted results of the \lya\ LF evolution may be caused by small statistics 
and systematic uncertainties such as contamination and cosmic variance.
To reliably investigate the evolution of the \lya\ LF at $\zi \gtrsim 7$,
one needs an ultra-deep narrowband imaging survey in large areas 
down to the \lya\ luminosity limit comparable to those of $\zi \le 6.6$ LAE samples.

Studies of the \lya\ LF evolution are important for understanding galaxy evolution and cosmic reionization.
The \lya\ damping wing of neutral hydrogen in inter-galactic medium (IGM) around galaxies attenuates \lya\ photons significantly.
Thus, a volume-averaged neutral hydrogen fraction, \xhi, of IGM would be constrained by
the evolution of the \lya\ LF at the epoch of \xhi\ $\sim 0.1 - 1.0$
\citep{2004ApJ...617L...5M, 2005pgqa.conf..363H, 2006Natur.443..186I, 2006ApJ...648....7K, 
2011ApJ...734..119K, 2008ApJ...677...12O, 2010ApJ...722..803O, 2010ApJ...723..869O, 2012ApJ...752..114S}.
The evolution of the \lya\ luminosity density between $\zi = 5.7$ and $6.6$ suggests
\xhi\ $= 0.2 \pm 0.2$ at $\zi = 6.6$ that is corrected for
the intrinsic UV luminosity evolution effect with the cosmic star formation rate density change \citep{2010ApJ...723..869O}.
A \lya\ emitting fraction of UV-continuum selected galaxies is similarly used for a probe of cosmic reionization.
Previous studies report that the \lya\ emitting fraction of Lyman break galaxies (LBGs)
decreases from $\zi \sim 6$ to $7$ in contrast to the increase of the \lya\ emitting fraction from $\zi \sim 3$ to $6$,
and claim that the neutral hydrogen fraction increases from $\zi \sim  6$ to $7$.
\citep{2011ApJ...743..132P, 2012ApJ...744..179S, 2012ApJ...744...83O, 2012ApJ...747...27T, 2012MNRAS.427.3055C, 2014MNRAS.443.2831C, 2014arXiv1403.5466P, 2014arXiv1404.4632S}.
By the comparison with theoretical models, these studies suggest \xhi\ $\gtrsim 0.5$ at $z\sim 7$.

There are other observational studies to investigate when and how cosmic reionization took place. 
Observations of the \cite{1965ApJ...142.1633G} (GP) trough in quasar (QSO) spectra indicate
\xhi\ $\sim 10^{-4}$ at $\zi \sim 6$ \citep{2006AJ....132..117F},
suggesting that cosmic reionization has been completed at this redshift.
Measurements of the polarization of cosmic microwave background (CMB) by \textit{WMAP}
constraint the optical depth of Thomson scattering, \tel\ $= 0.081 \pm 0.012$, and
indicate that the universe would be reionized at $\zi^\mathrm{inst}_\mathrm{re} = 10.1 \pm 1.0$ 
for the case of instantaneous reionization \citep{2013ApJS..208...19H, 2013ApJS..208...20B}.
Recent observational studies with \textit{Planck} show that the electron scattering optical depth is \tel\ $= 0.089^{+0.012}_{-0.014}$, 
and that the instantaneous reionization redshift is $\zi^\mathrm{inst}_\mathrm{re} = 11.1 \pm 1.1$ \citep{2013arXiv1303.5076P}.
\cite{2006PASJ...58..485T} estimate \xhi\ with the shape of \lya\ damping wing 
absorption found in the optical afterglow spectrum of GRB 050904 at $\zi \sim 6.3$,
and obtain \xhi\ $< 0.17$ ($68\%$ confidence level) at this redshift.
In the recent study with a GRB, the unprecedentedly bright optical afterglow spectrum of GRB 130606A at $\zi \sim 5.9$ 
suggests \xhi\ $=0.1- 0.5$ \citep{2014PASJ...66...63T}.
\cite{2011Natur.474..616M} report observations of a QSO at $\zi = 7.085$, ULAS J1120$+$0641,
and claim \xhi\ $> 0.1$ at this redshift from the near-zone transmission profile.
\cite{2011MNRAS.416L..70B} use radiative transfer simulations
with model absorptions of inhomogeneous IGM around ULAS J11201$+$0641,
and obtain \xhi\ $\gtrsim 0.1$.

Although the CMB observations rule out the instantaneous reionization at a late epoch,
it is difficult to understand how the reionization proceeds in the cosmic history.
As illustrated in Figure 23 of \cite{2010ApJ...723..869O},
there are large uncertainties of the \xhi\ estimates from the previous observational studies,
and one cannot distinguish between various models of reionization history.
A redshift of $\sim 7$ is the observational limit of optical instruments that enable us to conduct a deep and wide-field imaging survey.
The differences of reionization history in models are relatively large at $\zi \gtrsim 7$
\citep[see Figure 23 of ][]{2010ApJ...723..869O}. 
A measurement of the \lya\ LF at $z\sim 7$ with a good statistical accuracy is
useful to constrain \xhi\ near the observational limit and to address this issue of cosmic reionization history.

In this paper, we present the results of our ultra-deep narrowband imaging survey for $\zi = 7.3$ LAEs.
Using this sample, we derive the \lya\ LF with accuracies significantly better than those of previous $\zi \gtrsim 7$ studies.
We investigate the \lya\ LF evolution at $\zi \gtrsim 7$ with this \lya\ LF, and discuss the cosmic reionization history.
We describe the details of our $\zi = 7.3$ LAE survey and selection of our LAE candidates in Section \ref{sec:observation}.
We derive the $\zi = 7.3$ \lya\ LF and compare it with previous studies of $\zi \simeq 7.3$ in Section \ref{sec:LyaLF}. 
We examine the evolution of the \lya\ LF at $\zi = 5.7 - 7.3$,
and discuss cosmic reionization with the constraints of the electron scattering optical depth measurements of CMB in Section \ref{sec:discuss}.
Throughout this paper, we adopt AB magnitudes \citep{1974ApJS...27...21O}
and concordance cosmology with a parameter set of 
($h$, $\Omega_\mathrm{m}$, $\Omega_\Lambda$, $\sigma_8$) = (0.7, 0.3, 0.7, 0.8) consistent 
with the \textit{WMAP} and \textit{Planck} results \citep{2013ApJS..208...19H, 2013arXiv1303.5076P}.

\begin{figure}
\centering
\includegraphics[width=8cm]{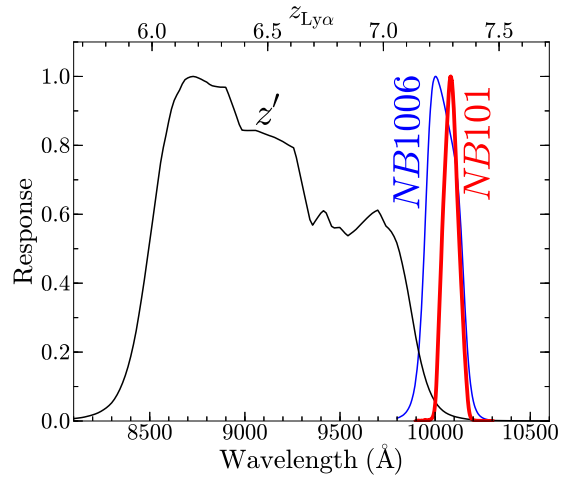}
\caption{Filter response curve of \textit{NB101} that is shown with the red line.
The blue and black lines represent the response curves of \textit{NB1006} and
\textit{z}$'$ bands, respectively. These response curves 
are based on actual lab measurements, and
include the quantum efficiency of Hamamatsu CCDs 
\citep{2008SPIE.7021E..52K}, airmass, transmission$+$reflection of instrument, and telescope optics.
For clarity, peaks of these curves are normalized to 1.0.
The upper abscissa axis indicates a redshift of \lya\ that corresponds to 
the wavelength. Note that \textit{NB1006} widely covers a redshift range of $z=7.2-7.3$,
and that \textit{NB101} targets a narrow redshift range centered at $z=7.3$.
} 
\label{fig:filter_res}
\end{figure}

\begin{deluxetable*}{lcccccc}
\tabletypesize{\footnotesize}
\tablecaption{Summary of Our Observations and Data\label{table:imaging_data}}
\tablewidth{0pt}
\tablehead{
\colhead{Field} & 
\colhead{Band} &
\colhead{Exposure Time} &
\colhead{PSF size\tablenotemark{a}} &
\colhead{Area} &
\colhead{$m_\mathrm{lim}$\tablenotemark{b}} &
\colhead{Date of Observations}\\
\colhead{} & 
\colhead{} & 
\colhead{(s)} &
\colhead{(arcsec)} &
\colhead{(arcmin$^2$)} & 
\colhead{($5\sigma$ AB mag)} &
\colhead{}  
}
\startdata
SXDS	& \textit{NB101}		& 57600					& 0.78		& 				& 24.6		& 2010 Dec 29$-$2011 Jan 1	\\
		& \textit{NB101}		& 73200					& 0.86		& 				& 24.3		& 2012 Dec 11$-$14		\\
		& \textit{NB101} (Final)	& 130800\tablenotemark{c}	& 0.80		& $7.9 \times 10^2$	& 24.9		& ---						\\
\addlinespace[2pt]
\hline
\addlinespace[2pt]
COSMOS	& \textit{NB101}		& 83510.6					& 0.72		& 				& 24.8		& 2010 Dec 29$-$2011 Jan 2	\\
		& \textit{NB101}		& 61200					& 0.99		& 				& 23.7		& 2012 Dec 11$-$14		\\
		& \textit{NB101}		& 105600					& 0.90		&				& 24.4		& 2013 Feb 9$-$12			\\
		& \textit{NB101} (Final)	& 250310.6\tablenotemark{c}	& 0.77		& $8.4 \times 10^2$	& 25.1		& ---						\\
\addlinespace[2pt]
\hline
\hline
\addlinespace[2pt]
\multicolumn{7}{c}{Archival Broadband Data\tablenotemark{d}}	\\
\addlinespace[2pt]
\hline
\addlinespace[2pt]
SXDS	& \textit{B}			& 						& 0.84		& $7.9 \times 10^2$	& 28.1		&						\\
		& \textit{V}			& 						& 0.84		&				& 27.7		&						\\
		& \textit{R}			& 						& 0.84		&				& 27.6		&						\\
		& \textit{i}$'$		& 						& 0.84		&				& 27.3		&						\\
		& \textit{z}$'$		& 						& 0.80		&				& 26.9		&						\\
\addlinespace[2pt]
\hline
\addlinespace[2pt]
COSMOS	& \textit{B}			& 						& 0.95		& $8.4 \times 10^2$	& 27.7		&						\\
		& \textit{V}			& 						& 1.32		&				& 26.4		&						\\
		& \textit{R}			& 						& 1.05		&				& 26.9		&						\\
		& \textit{i}$'$		& 						& 0.95		&				& 26.6		&						\\
		& \textit{z}$'$		& 						& 0.84		&				& 26.8		&											
\enddata
\tablenotetext{a}{The FWHM value of PSF.}
\tablenotetext{b}{The $5\sigma$ limiting magnitude in a circular aperture with a diameter of $2\times$ PSF FWHM.}
\tablenotetext{c}{The total on-source integration times correspond to 36.3 and 69.5 hours in the SXDS and COSMOS fields, respectively.}
\tablenotetext{d}{The Broadband images are archival data presented in \cite{2008ApJS..176....1F} for SXDS and \cite{2007ApJS..172...99C} for COSMOS.
The SXDS and COSMOS \textit{z}$'$ data include the images taken by the Subaru intensive program conducted in $2009 - 2011$ (PI: H. Furusawa).}
\end{deluxetable*}

\section{Imaging Observations and\\
Data Reduction}\label{sec:observation}

\subsection{NB101 Observations} \label{sec:nb101obs}

We have carried out an ultra-deep large-area narrowband imaging survey with Subaru/Suprime-Cam
\citep{2002PASJ...54..833M} to study LAEs at $z=7.3$ down to the faint \lya\ luminosity limit.
For these observations, we have developed a new custom narrowband filter, \textit{NB101}.
The filter transmission of \textit{NB101} is centered at $\lambda_\mathrm{c} = 10095 \mathrm{\AA}$ and 
\textit{NB101} is designed to have a narrow and sharp FWHM of $\Delta \lambda = 90 \mathrm{\AA}$.
The \textit{NB101} filter identifies LAEs in the redshift range of $\zi = 7.302 \pm 0.037$.
We show the filter response curve of our \textit{NB101} in Figure \ref{fig:filter_res}.
Note that there is a Suprime-Cam narrowband filter,
\textit{NB1006}, at a similar wavelength \citep{2012ApJ...752..114S}. 
The \textit{NB1006} filter has a central wavelength of $\lambda_\mathrm{c} = 10052 \mathrm{\AA}$
slightly bluer than that of our \textit{NB101} and an FWHM of $\Delta \lambda = 214 \mathrm{\AA}$
about 2-3 times broader than that of our \textit{NB101}.
Similarly, there is another Suprime-Cam narrowband filter of \textit{NB973} targeting $z=7.0$ LAEs
with a central wavelength of $\lambda_\mathrm{c} = 9755 \mathrm{\AA}$
and an FWHM of $\Delta \lambda \sim 200 \mathrm{\AA}$ \citep{2006Natur.443..186I, 2008ApJ...677...12O, 2010ApJ...722..803O}
that is also much broader than the FWHM of our \textit{NB101} filter.
Since our \textit{NB101} filter has a significantly narrower/shaper FWHM than the \textit{NB1006} and \textit{NB973} filters,
our \textit{NB101} filter is more sensitive to an emission line than the \textit{NB1006} and \textit{NB973} filters.
Although the survey volume is smaller for \textit{NB101} than for \textit{NB1006} and \textit{NB973},
the line sensitivity is more important for the observational studies of $z\gtrsim 7$ sources whose
LF's exponential edge is near the observational limit.
At $z=7.3$, our \textit{NB101} filter allows us to reach a \lya\ flux limit faster 
than the previous \citeauthor{2012ApJ...752..114S}'s \textit{NB1006} surveys by $\sim 160$\%.
Thus, we can reach the \lya\ luminosity limit of LAEs much fainter than the previous Subaru studies
for $\zi \sim 7$ LAEs.

With our new \textit{NB101} filter,
we observed two independent fields, Subaru/\textit{XMM-Newton} Deep Survey (SXDS) and Cosmic Evolution Survey (COSMOS) fields.
The SXDS field is located at $02^\mathrm{h} 18^\mathrm{m} 00^\mathrm{s}.0$, $-05^\mathrm{d} 00' 00''$ (J2000) 
\citep{2008ApJS..176....1F, 2008ApJS..176..301O, 2010ApJ...723..869O}, 
It consists of five subfields of $\sim 0.2$ deg$^2$, SXDS-C, N, S, E and W.
We choose a field of $\sim 0.2$ deg$^2$ to cover the southern half of SXDS-C and the northern half of SXDS-S, 
where bright stars do not exist and \textit{HST} CANDELS \citep{2011ApJS..197...35G, 2011ApJS..197...36K}, 
\textit{UKIRT} UKIDSS \citep{2007MNRAS.379.1599L}, 
\textit{Spitzer} SpUDS (PI: J. Dunlop), and SEDS (PI: G. Fazio) data are also available \citep[see Figure 2 in][]{2010ApJ...722..803O}.
The target field of COSMOS is an area of $\sim 0.2$ deg$^2$
centered at $10^\mathrm{h} 00^\mathrm{m} 28^\mathrm{s}.6$, $+02^\mathrm{d} 12' 21''.0$ (J2000) \citep{2007ApJS..172....1S}.
In the COSMOS field, there exist CANDELS and UltraVISTA (PI: J. Dunlop) imaging data.
Each of SXDS and COSMOS field is covered by one pointing of Suprime-Cam whose field of view is 
918 arcmin$^2$.
Our observations were conducted in $2010-2013$.
The total on-source integration time is 106 hours where 
36.3 and 69.5-hour data were obtained in the SXDS and COSMOS fields, respectively.
We summarize the details of our observations as well as image qualities in Table \ref{table:imaging_data}.
In addition to these \textit{NB101} images, we use archival data of deep broadband (\textit{B}, \textit{V}, \textit{R}, \textit{i}$'$ and \textit{z}$'$) 
images of the SXDS and COSMOS projects 
(\citealt{2008ApJS..176....1F} and \citealt{2007ApJS..172...99C} for SXDS and COSMOS fields, respectively).
The properties of these broadband-data are also listed in Table \ref{table:imaging_data}.

Our \textit{NB101} data are reduced with the Suprime-Cam Deep field REDuction package \citep[SDFRED;][]{2002AJ....123...66Y, 2004ApJ...611..660O}.
In this reduction process, we perform bias subtraction, flat-fielding, distortion+atmospheric-dispersion correction, 
sky subtraction, image alignments, and stacking.
Before the image alignments, we mask out areas contaminated by spurious signals and meteor+satellite trails.
We remove cosmic rays with a rejected-mean algorithm.
We make composite images from each one-night data set.
Then, we stack all of these one-night composite images with weights based on signal-to-noise ratios of these images to make the final images.
To obtain the weights, we measure the photometric-zero point and the limiting magnitude
for each one-night composite image.
For measuring colors of objects precisely, we align the reduced \textit{NB101} images with the broadband images 
based on hundreds of bright stellar objects commonly detected in the \textit{NB101} and broadband images.
We calculate the photometric-zero points from our standard star data.
Estimating the photometric-zero points, we took data of spectrophotometric standard stars of 
G191-B2B and GD153 \citep{1995AJ....110.1316B} with \textit{NB101}.
We check these photometric-zero points with
stellar sequences of observed stellar objects in our fields and 
175 Galactic stars of \cite{1983ApJS...52..121G}
on a two-color diagram of \textit{z}$'$ $-$ \textit{NB101}
vs. \textit{i}$'$ $-$ \textit{z}$'$.
After the check of the photometric-zero points,
we estimate the limiting magnitudes of our images.

The final \textit{NB101} images of SXDS and COSMOS have 
the seeing size of $\simeq 0.''8$, and reach the 5$\sigma$ magnitude of $\simeq 25.0$ mag.
We summarize the qualities of our final \textit{NB101} images in Table \ref{table:imaging_data}.
We use pixels of the imaging data neither contaminated with halos of bright stars, CCD blooming, nor low signal-to-noise ratio region
near the edge of Suprime-Cam field of view. These low-quality regions are masked out, and
the effective survey areas are $7.9 \times 10^2$ and $8.4 \times 10^2 \ \mathrm{arcmin}^2$
in the SXDS and COSMOS fields, respectively.
Thus, our total survey area is $1.6 \times 10^3 \ \mathrm{arcmin}^2$, i.e. $\simeq 0.5$ deg$^2$.
If we assume a simple top-hat selection function for LAEs whose redshift distribution is defined by the FWHM of \textit{NB101},
these effective survey areas correspond to the comoving
survey volumes of $1.2 \times 10^5$ and $1.3 \times 10^5 \ \mathrm{Mpc}^3$
for the SXDS and COSMOS fields, respectively.

\subsection{Photometry} \label{sec:photo}

The source detection and photometry are performed with SExtractor version 2.5.0 \citep{1996A&AS..117..393B}.
Sources are identified with the criterion: contiguous $>5$ pixels with
a flux greater than the $2\sigma$ level of sky fluctuation.
We conduct the source detection in our \textit{NB101} images, and 
obtain the broadband photometry at the positions of the sources.
We detect a total of 69,387 objects in the SXDS and COSMOS fields
down to the $5\sigma$ limits of aperture magnitudes
that are \textit{NB101} $= 24.9$ (SXDS) and $25.1$ (COSMOS).
Here, we define the aperture magnitude of \texttt{MAG\_APER} of SExtractor
with an aperture size of $2\times$ PSF FWHM,
and use the aperture magnitude for measuring
colors of objects.
For total magnitude estimates,
we apply an aperture correction value of $0.3$ to the aperture magnitudes.
Because \texttt{MAG\_AUTO} of SExtractor gives biased magnitude measurements
for faint objects around the detection limits, we use this aperture correction technique.
\cite{2012ApJ...744...83O} study \textit{z}-dropout galaxies at $\zi \sim 7$ using Subaru/Suprime-Cam data, 
and derive the aperture correction value of $\sim 0.3$ mag.
We apply the same aperture correction value as \cite{2012ApJ...744...83O}
because the PSF FWHM of the Ono et al's data ($\sim 0.''8 - 0.''9$) is similar to that of our NB101 data.
The reliability of this technique is investigated in Section \ref{sec:z7p3_cand}.

\subsection{Photometric Sample of $\zi\ =7.3$ LAEs} \label{sec:z7p3_cand}

\begin{figure}
\centering
\includegraphics[width=8cm]{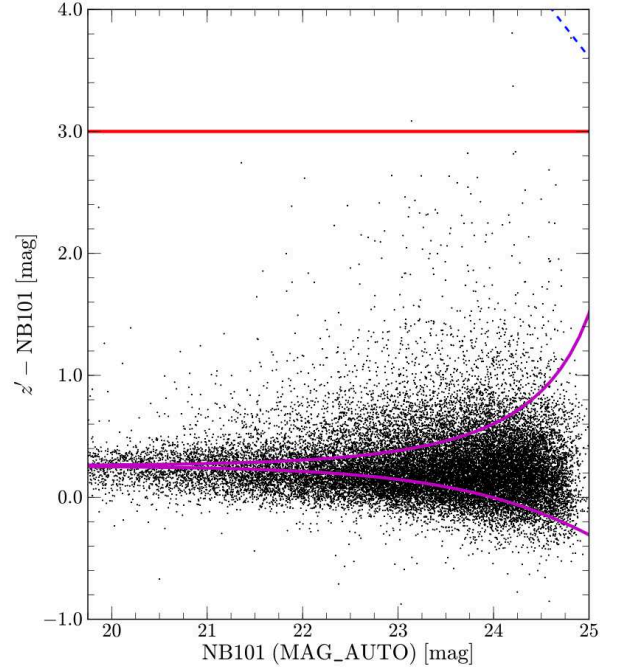}
\caption{Color-magnitude diagram of $\textit{z}' - \textit{NB101}$ vs. \textit{NB101}
for objects detected in the SXDS field.
The black dots show all of the detected objects.
The blue-dashed and magenta-solid lines indicate the $1\sigma$ limit of the $\textit{z}'$ magnitude
and the $3\sigma$ error of the $\textit{z}' - \textit{NB101}$ color, respectively.
The red line represents the $\textit{z}' - \textit{NB101}$ color criterion for the selection of our $\zi = 7.3$ LAEs.
} 
\label{fig:CMD_sxds}
\end{figure}

\begin{figure}
\centering
\includegraphics[width=8cm]{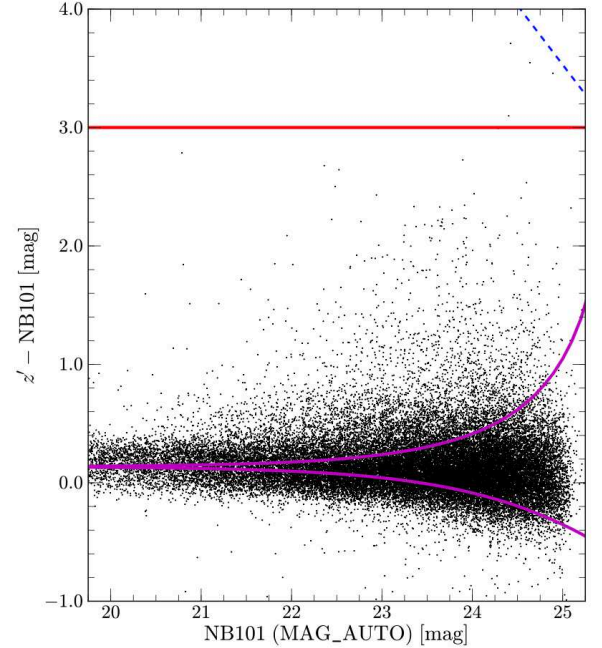}
\caption{Same as Figure \ref{fig:CMD_sxds}, but for the COSMOS field.
} 
\label{fig:CMD_cosmos}
\end{figure}

We isolate $\zi = 7.3$ LAE candidates from all of the objects detected in Section \ref{sec:photo}
based on a narrowband excess of \lya\ emission and no detection of blue continuum flux.
Figures \ref{fig:CMD_sxds} and \ref{fig:CMD_cosmos} show 
the color-magnitude diagrams of the \textit{NB101} magnitude and
the narrowband excess color, $\textit{z}' - $ \textit{NB101}, 
for the objects detected in SXDS and COSMOS fields.
The detected objects have a color of $\textit{z}' - $\textit{NB101} $\simeq +0.2$ on average
in the magnitude range of $22 < $ \textit{NB101} $< 24$.
To determine the $\textit{z}' - $ \textit{NB101} color criterion for our LAE candidate selection,
we assume a model spectrum of $\zi = 7.3$ LAE that has a \lya\ line and 
a flat ultraviolet (UV) continuum (i.e., $f_{\nu} =$ const.) with an IGM absorption \citep{1995ApJ...441...18M}.
Based on the model spectrum, we adopt the criterion that $\textit{z}' -  \textit{NB101} \geqslant 3.0$,
which corresponds to LAEs with the rest-frame equivalent width, EW$_0$, 
of $\mathrm{EW}_{0} \gtrsim 0 \mathrm{\AA}$, which is similar to the criterion adopted by
\cite{2012ApJ...752..114S}. 
Note that this small limit of the EW$_0$ criterion gives a chance to select 
high-\zi\ dropout galaxies and foreground red objects,
due to photometric errors, which are the potential contamination sources.
Because this EW$_0$ limit gives a more complete sample, we apply this EW$_0$ limit.
We discuss the effect of this small EW$_0$ limit in Section \ref{sec:evolution}.

Adding other criterion of no detectable continuum flux bluer than \lya ,
we define the selection criteria of $\zi = 7.3$ LAEs:
\begin{align}
&\textit{NB101} < \textit{NB101}_{5\sigma} \ \mathrm{and} \ \textit{B} > \textit{B}_{3\sigma} 	\notag	\\
&\mathrm{and} \ \textit{V} > \textit{V}_{3\sigma} \ \mathrm{and} \ \textit{R} > \textit{R}_{3\sigma} \ \mathrm{and} \ \textit{i}' > \textit{i}'_{3\sigma}	\label{eq:criteria}	\\
&\mathrm{and} \ \left[ (\textit{z}' - \textit{NB101} \geqslant 3.0 ) \ \mathrm{or} \ ( \textit{z}' > \textit{z}'_{3\sigma} )\right],	\notag	
\end{align}
where the indices of 5$\sigma$ and 3$\sigma$ denote the 5$\sigma$ and 3$\sigma$ detection limits of the images, respectively.

\begin{figure*}
\centering
\includegraphics[width=0.9\textwidth]{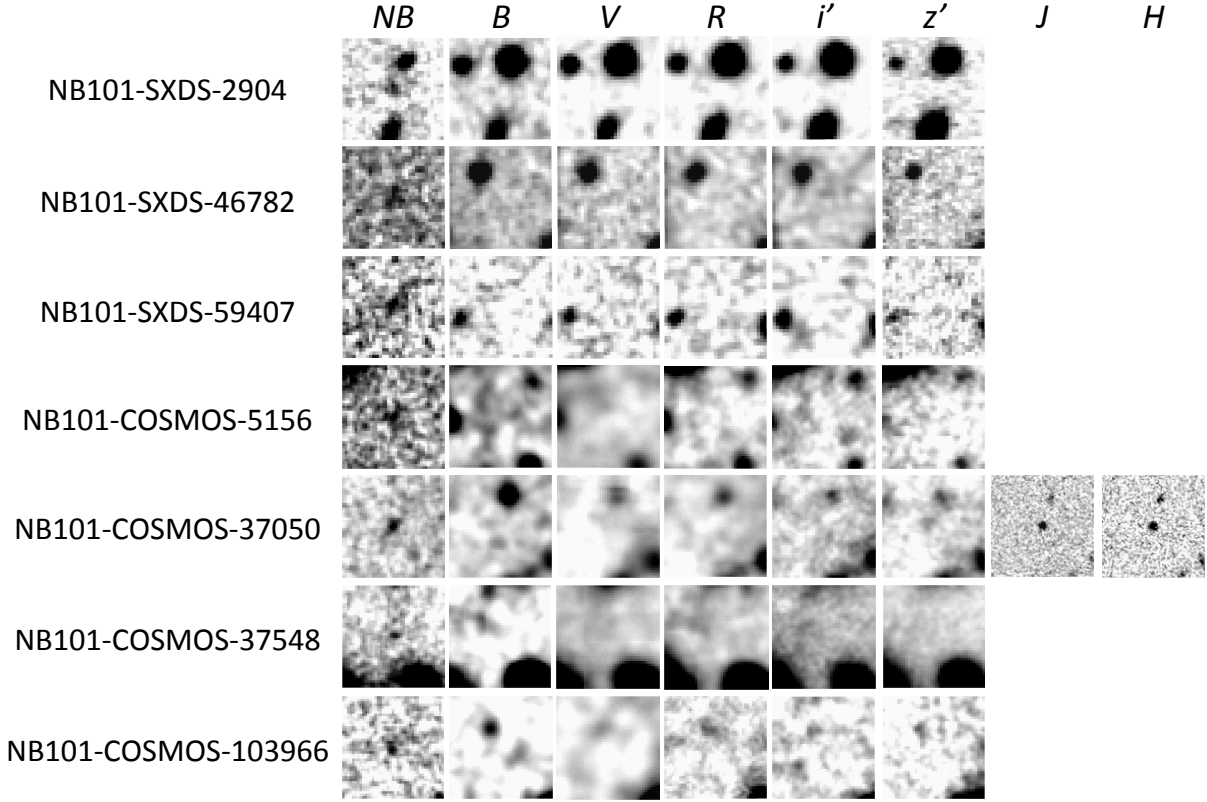}
\caption{Snapshots of our $\zi = 7.3$ LAE candidates. 
The size of each image is $6'' \times 6''$.
North is up and east is to the left.
} 
\label{fig:snapshot}
\end{figure*}

We apply these photometric criteria to all of our detected objects, and
identify three and four $\zi = 7.3$ LAE candidates in the SXDS and COSMOS fields, respectively.
We show the snapshot images of these LAE candidates in Figure \ref{fig:snapshot}.
In our \textit{NB101} ultra-deep survey, we have reached a $5\sigma$ limiting flux of
$\simeq 6.5 \times 10^{-18}$ erg s$^{-1}$ cm$^{-2}$ corresponding to a limiting luminosity of 
$L_{\mathrm{Ly}\alpha} \simeq 4.1 \times 10^{42}$ erg s$^{-1}$ in the SXDS field,
and $\simeq 3.8 \times 10^{-18}$ erg s$^{-1}$ cm$^{-2}$ equivalent to
$L_{\mathrm{Ly}\alpha} \simeq 2.4 \times 10^{42}$ erg s$^{-1}$ in the COSMOS field.
These limiting luminosities are derived from the $5\sigma$ limiting magnitudes of \textit{NB101}
and the $1\sigma$ limiting magnitudes of $\textit{z}'$.
These are conservative estimates, because the limiting luminosity values are 
larger than those calculated with the $> 1\sigma$ limiting magnitudes of $\textit{z}'$.
In the calculations for the \lya\ luminosities, we assume that \lya\ emission is placed at the central wavelength of the narrow band.
\cite{2008ApJS..176..301O} derive the \lya\ luminosities in the same manner as ours,
and compare these \lya\ luminosities with spectroscopic luminosities.
They find that both measurements agree well within the error bars, and that the difference of these values is small
(see Section \ref{sec:z7p3lyalf} for the effects of the \lya\ luminosity uncertainties in \lya\ LF derivation).
The \textit{NB101} image of COSMOS field is the deepest image in our \textit{NB101} data.
The $5\sigma$ limiting luminosity in the COSMOS field is about 4 times deeper than previous Subaru 
studies for LAEs at $\zi \sim 7$ \citep{2008ApJ...677...12O, 2010ApJ...722..803O, 2012ApJ...752..114S}.
Moreover, the $5\sigma$ limiting luminosity is comparable with those of previous Subaru $\zi = 3.1-6.6$ LAE surveys
\citep{2006PASJ...58..313S, 2006ApJ...648....7K, 2011ApJ...734..119K, 2008ApJS..176..301O, 2010ApJ...723..869O}.

We present the photometric properties of our $\zi = 7.3$ LAE candidates in Table \ref{table:z7p3candidate}.
The total magnitudes listed in Table \ref{table:z7p3candidate} are obtained by the aperture-correction technique
explained in Section \ref{sec:photo}. We compare the total magnitude and \texttt{MAG\_AUTO} of SExtractor
for the most luminous LAE candidate that probably includes a negligible bias in the \texttt{MAG\_AUTO} estimate,
and find that these two magnitudes are consistent within the errors. Thus, it is reasonable to use the total magnitudes
given by the aperture-correction technique that requires the assumption that $\zi = 7.3$ LAEs are point sources
\citep{2010ApJ...724.1524O, 2012ApJ...744...83O}.

We investigate our $\zi = 7.3$ LAE candidates in the \textit{J} and \textit{H} images of \textit{HST} CANDELS fields 
that are subfields of our COSMOS and SXDS survey areas.
Two out of seven LAE candidates, NB101-COSMOS-37050 and NB101-COSMOS-37548, fall in the CANDELS field of COSMOS.
We detect NB101-COSMOS-37050 both in the \textit{J} and \textit{H} images (Figure \ref{fig:snapshot}), 
but NB101-COSMOS-37548 neither in the \textit{J} nor \textit{H} data.
We obtain the \textit{J} and \textit{H} magnitudes of NB101-COSMOS-37050, and present the magnitudes in Table \ref{table:z7p3candidate}.
There are no counterparts of LAE candidates found in the CANDELS field of SXDS.

We examine whether $\zi = 7.2-7.3$ LAEs found by \cite{2012ApJ...752..114S} are identified in our \textit{NB101} data.
\cite{2012ApJ...752..114S} have observed the SXDS subfield same as our survey area 
with their \textit{NB1006} filter (Section \ref{sec:nb101obs}),
and obtained two photometric LAE candidates, SXDS-NB1006-1 and SXDS-NB1006-2.
However, both of two LAEs of \cite{2012ApJ...752..114S} are not detected 
in our \textit{NB101} images.  Because \cite{2012ApJ...752..114S} report that their spectroscopy indicates
that one of them, SXDS-NB1006-2, resides at $\zi = 7.215$, the redshift of SXDS-NB1006-2 is 
out of our survey redshift range of $\zi = 7.302 \pm 0.037$
where \textit{NB101} has a sensitivity for a Ly$\alpha$ emission line.
Thus, the reason for no detection of SXDS-NB1006-2 
is clear, while the reason for another object, SXDS-NB1006-1, is unknown. 
Since SXDS-NB1006-1 is not confirmed by their spectroscopic follow-up observations,
it is possible that the \lya\ emission of SXDS-NB1006-1 also falls in the wavelength
where \textit{NB101} does not cover. Note that the FWHM of \textit{NB1006} is $214$\AA,
while \textit{NB101} is only $90$\AA\ (Section \ref{sec:nb101obs}).

\begin{deluxetable*}{lcccccccccc}
\tabletypesize{\footnotesize}
\tablecaption{Our $\zi = 7.3$ LAE Candidates\label{table:z7p3candidate}}
\tablewidth{0pt}
\tablehead{
\colhead{ID} & 
\colhead{\textit{B}} &
\colhead{\textit{V}} &
\colhead{\textit{R}} &
\colhead{$\textit{i}'$} &
\colhead{$\textit{z}'$} &
\colhead{\textit{NB101}\tablenotemark{a}}	&
\colhead{\textit{NB101}(total)\tablenotemark{b}}	&
\colhead{\textit{J}\tablenotemark{a}} &
\colhead{\textit{H}\tablenotemark{a}}	 &
\colhead{$L(\mathrm{Ly}\alpha)$}	\\
\colhead{} & 
\colhead{} &
\colhead{} &
\colhead{} &
\colhead{} &
\colhead{} &
\colhead{}	&
\colhead{}	&
\colhead{} &
\colhead{} &
\colhead{($10^{42}$ erg s$^{-1}$)}
}
\startdata
NB101-SXDS-2904			& $>28.6$	& $>28.3$	& $>28.1$	& $>27.8$	& $>27.4$	& $24.50^{+0.16}_{-0.14}$& $24.20^{+0.12}_{-0.11}$& ---	& ---	& 9.68 \\
NB101-SXDS-46782		& $>28.6$	& $>28.3$	& $>28.1$	& $>27.8$	& $>27.4$	& $24.84^{+0.23}_{-0.19}$& $24.54^{+0.17}_{-0.15}$& ---	& ---	& 5.72 \\
NB101-SXDS-59407		& $>28.6$	& $>28.3$	& $>28.1$	& $>27.8$	& $>27.4$	& $24.80^{+0.22}_{-0.18}$& $24.50^{+0.16}_{-0.14}$& ---	& ---	& 6.13 \\
\addlinespace[2pt]
\hline
\addlinespace[2pt]
NB101-COSMOS-5156		& $>28.3$	& $>27.0$	& $>27.4$	& $>27.2$	& $>27.3$	& $24.98^{+0.21}_{-0.18}$& $24.68^{+0.16}_{-0.14}$& ---	& ---	& 3.82 \\
NB101-COSMOS-37050		& $>28.3$	& $>27.0$	& $>27.4$	& $>27.2$	& $>27.3$	& $24.84^{+0.19}_{-0.16}$& $24.54^{+0.14}_{-0.12}$& $25.42^{+0.05}_{-0.05}$& $25.39^{+0.06}_{-0.05}$	& 5.11 \\
NB101-COSMOS-37548		& $>28.3$	& $>27.0$	& $>27.4$	& $>27.2$	& $>27.3$	& $25.03^{+0.23}_{-0.19}$& $24.73^{+0.17}_{-0.14}$& ---	& ---	& 3.39 \\
NB101-COSMOS-103966	& $>28.3$	& $>27.0$	& $>27.4$	& $>27.2$	& $>27.3$	& $25.07^{+0.23}_{-0.19}$& $24.77^{+0.17}_{-0.15}$& ---	& ---	& 3.01 
\enddata
\tablenotetext{a}{The magnitudes with the $1\sigma$ error measured with an aperture whose diameter is $2\times$ PSF FWHM.}
\tablenotetext{b}{The total magnitudes which are obtained by the aperture-correction technique explained in Section \ref{sec:photo}}.
\end{deluxetable*}

\begin{deluxetable*}{lccc}
\tabletypesize{\footnotesize}
\tablecaption{Magnitudes of our $\zi = 7.3$ LAE Candidates in the Different Epoch\label{table:z7p3variability}}
\tablewidth{0pt}
\tablehead{
\colhead{ID} & 
\colhead{\textit{NB101}\tablenotemark{a}} & 
\colhead{\textit{NB101}\tablenotemark{a}} & 
\colhead{\textit{NB101}\tablenotemark{a}}	\\
\colhead{} &
\colhead{(2010)\tablenotemark{b}} & 
\colhead{(2012)\tablenotemark{b}} & 
\colhead{(2013)\tablenotemark{b}} 
}
\startdata
NB101-SXDS-2904			&$24.51^{+0.22}_{-0.18}$& $24.38^{+0.26}_{-0.21}$& ---		\\
NB101-SXDS-46782		&$24.74^{+0.28}_{-0.22}$& $25.03^{+0.54}_{-0.36}$& ---		\\
NB101-SXDS-59407		&$24.79^{+0.30}_{-0.23}$& $24.76^{+0.40}_{-0.29}$& ---		\\
\addlinespace[2pt]
\hline
\addlinespace[2pt]
NB101-COSMOS-5156		&$25.03^{+0.31}_{-0.24}$& $24.65^{+0.71}_{-0.43}$& $24.86^{+0.40}_{-0.29}$	\\
NB101-COSMOS-37050		&$24.64^{+0.21}_{-0.17}$& $24.69^{+0.75}_{-0.44}$& $25.48^{+0.84}_{-0.47}$	\\
NB101-COSMOS-37548		&$25.19^{+0.37}_{-0.27}$& $24.07^{+0.36}_{-0.27}$& $24.92^{+0.42}_{-0.30}$	\\
NB101-COSMOS-103966	&$25.11^{+0.34}_{-0.26}$& $24.92^{+1.04}_{-0.52}$& $24.76^{+0.35}_{-0.27}$	
\enddata
\tablenotetext{a}{The magnitudes with the $1\sigma$ error measured with an aperture whose diameter is $2\times$ PSF FWHM.}
\tablenotetext{b}{The values in parenthesis present the epochs of data used for the stacked images. 2010, 2012, and 2013 indicate the epochs of $2010-2011$, $2012$, and $2013$ observing periods, respectively.
}
\end{deluxetable*}

\section{Luminosity Function} \label{sec:LyaLF}

\subsection{Contamination of Our Sample} \label{sec:contami}

We investigate the contamination of our $\zi = 7.3$ LAE sample.
The sources of possible contamination are spurious objects, transients, and 
foreground interlopers. First, 
our \textit{NB101} images of SXDS and COSMOS fields are taken 
in 2010$-$2012 and 2010$-$2013, respectively (see Table \ref{table:imaging_data}).
We stack \textit{NB101} data of SXDS field observed in 2010$-$2011 and 2012, and obtain \textit{NB101} images for the two epochs.
Similarly, we make three \textit{NB101} stacked images of COSMOS field at three epochs, 2010$-$2011, 2012, and 2013.
The $5\sigma$ limiting magnitudes of these epoch images are summarized in Table \ref{table:imaging_data}.
The results of independent photometry at the different epochs of our observations are shown in Table \ref{table:z7p3variability}.
All of the magnitudes of our LAE candidates in the multi-epochs are consistent within the $\simeq 95$\%-significance levels of the photometric errors.
We find no variable signatures of transients in our LAEs. Because our LAEs are selected from narrowband images
taken over 3-4 years, a fraction of transient contamination in our LAE sample is very small.
Similarly, all of our LAEs, except NB101-SXDS-46782 and NB101-COSMOS-37050, 
are detected at the $>3$ sigma levels in the $\ge 2$ epoch images. Thus, our LAEs, except NB101-SXDS-46782
and NB101-COSMOS-37050, are not spurious sources. NB101-SXDS-46782 and NB101-COSMOS-37050 
are found only at the $\simeq 2\sigma$ levels in the 2012 and 2012-2013 epoch images, respectively.
However, we have identified the sources of NB101-SXDS-46782 and NB101-COSMOS-37050 
in these epoch images by visual inspection.
It is likely that NB101-SXDS-46782 and NB101-COSMOS-37050 are also not spurious sources.
Second, spectroscopic follow-up observations for one of our candidates, NB101-SXDS-2904, were conducted
with Keck/NIRSPEC, LRIS and MOSFIRE,
and a single emission line that is probably \lya\ is clearly detected 
from this object by all of these Keck spectroscopic observations (M. Ouchi et al. in preperation).
Although only one LAE in our sample is observed by spectroscopy,
no foreground interlopers are, so far, found by spectroscopic observations.

\subsection{Detection Completeness and Surface Number Density} \label{sec:detcomp_snd}

\begin{figure}
\centering
\includegraphics[width=8cm]{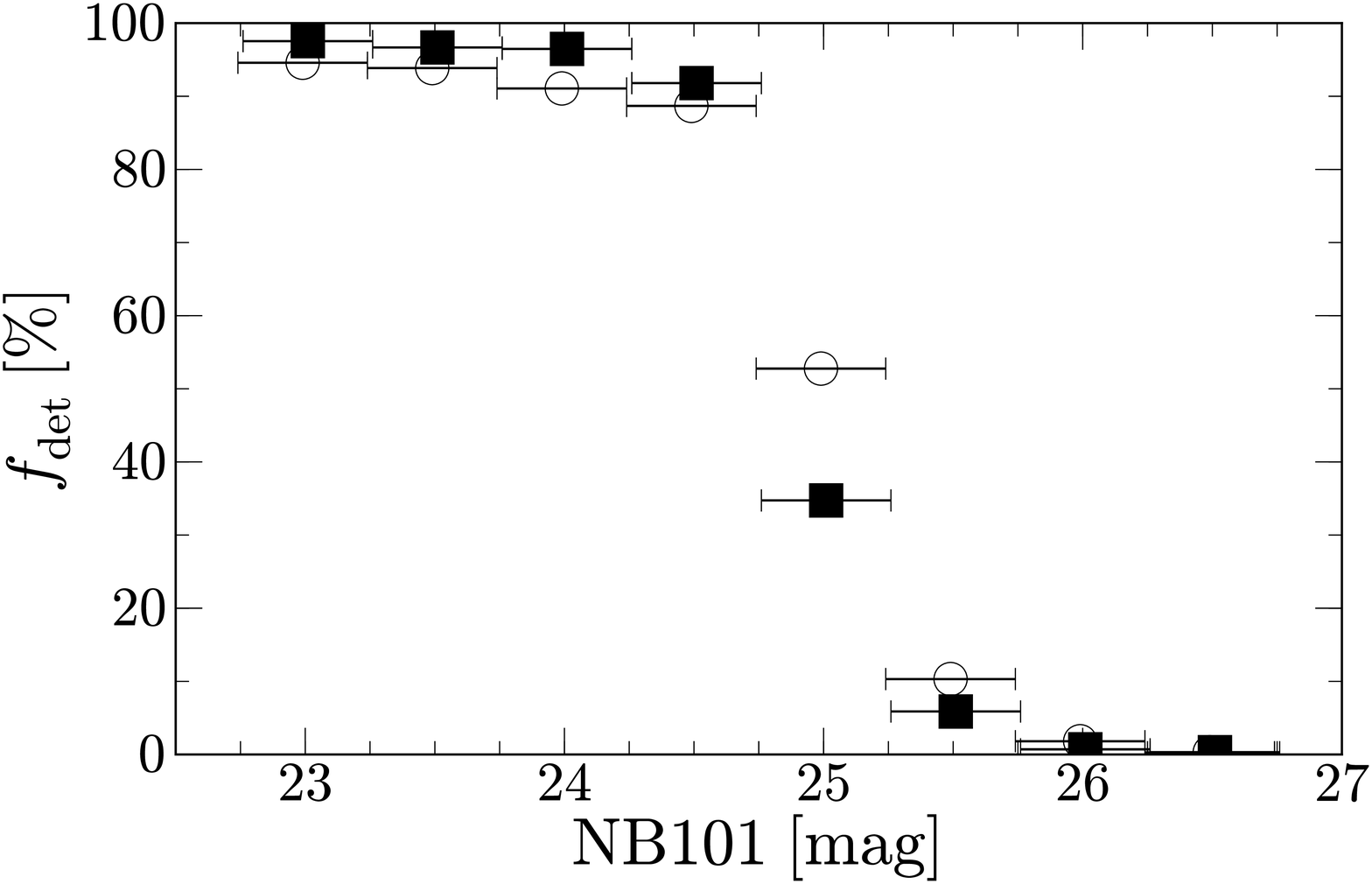}
\caption{Detection completeness of our \textit{NB101} images.
The filled squares and open circles represent the completeness in the SXDS and COSMOS fields, respectively.
} 
\label{fig:det_comp}
\end{figure}

We estimate detection completeness as a function of the \textit{NB101} magnitude by Monte-Carlo simulations.
We distribute a number of pseudo LAEs with various magnitudes in our \textit{NB101} images, and detect the pseudo LAEs 
in the same manner as our source extraction for real sources (Section \ref{sec:photo}).
Here, we assume that $\zi = 7.3$ LAEs are point sources whose profiles are obtained by
the stack of bright point sources in our \textit{NB101} images. We define the detection completeness
as the fraction of the numbers of the extracted pseudo LAEs to all of the input pseudo LAEs,
and obtain the detection completeness presented in Figure \ref{fig:det_comp}.
We find that the detection completeness is typically $\gtrsim 90 \%$ for luminous sources with \textit{NB101} $\lesssim 24.5$
and nearly $50 \%$ at around the $5\sigma$ limiting magnitude of \textit{NB101} $\simeq 25$.

\begin{figure}
\centering
\includegraphics[width=8cm]{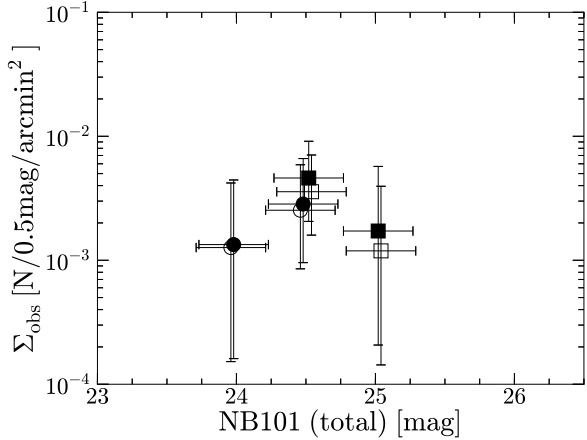}
\caption{Surface number density of our $\zi = 7.3$ LAEs as a function of the \textit{NB101} magnitude.
The circles and squares represent the surface number densities in the SXDS and COSMOS fields.
The filled and open symbols indicate the ones with and without the detection completeness correction, respectively.
} 
\label{fig:snd_all}
\end{figure}

Figure \ref{fig:snd_all} shows the surface number densities of $z=7.3$ LAEs.
The surface number densities are calculated by dividing the number counts of LAEs
by our effective survey areas shown in Section \ref{sec:nb101obs}.
We correct these surface number densities for the detection completeness.
The uncertainties of the surface densities of $\zi = 7.3$ LAEs are defined 
with the Poisson errors for small number statistics \citep{1986ApJ...303..336G}.
The values of columns ``0.8413'' in Tables 1 and 2 of \cite{1986ApJ...303..336G}
are used for the upper and lower limits of the Poisson errors, respectively.

\subsection{$\zi\ = 7.3$ \lya\ Luminosity Function} \label{sec:z7p3lyalf}

We calculate the \lya\ EW$_0$ of LAEs from the \textit{NB101} and $\textit{z}'$ magnitudes,
and estimate the \lya\ luminosities of LAEs with these EW$_0$ and the \textit{NB101} total magnitudes. 
For the errors of the \lya\ luminosities, we carry out Monte Carlo simulations under the assumption that
the spectrum of LAEs has a \lya\ line and a flat UV continuum (i.e., $f_{\nu} =$ const.) with the IGM absorption,
following the methods applied in \citet{2006PASJ...58..313S, 2008ApJS..176..301O, 2010ApJ...723..869O}.
Again, in the calculations for the \lya\ luminosities, we assume that \lya\ emission is placed at the central wavelength of the narrow band.
Similarly, we derive the \lya\ LF of $\zi =7.3$ LAEs in the same manner as \cite{2008ApJS..176..301O, 2010ApJ...723..869O}.
We calculate the volume number densities of LAEs in each \lya\ luminosity bin,
dividing the observed surface number densities of LAEs by our survey volumes
based on a top-hat filter transmission curve assumption.
This procedure of \lya\ LF derivation is known as the classical method.
Note that there are two uncertainties of the \lya\ LFs derived by the classical method.
(1) A \lya\ flux of a LAE at the fixed narrowband magnitude varies by the LAE's redshift.
(2) A redshift distribution of LAEs depends on a \lya\ EW.
In order to evaluate such uncertainties, \citet{2006PASJ...58..313S} and \cite{2008ApJS..176..301O} 
perform Monte Carlo simulations.
In these simulations, they generate a mock catalogue of LAEs with a set of Schechter parameters ($\phi^{*}$, $L^{*}$, $\alpha$)
and a Gaussian sigma for a probability distribution of \lya\ EW$_0$, and uniformly distribute the LAEs of the mock catalog
in a comoving volume over the redshift range that a narrowband covers.
They ``observe'' these LAEs with the narrow and broad bands to be the same as the real band response.
They select LAEs using the same criteria as was used for selecting the actual LAEs 
and derive the number densities and color distributions from the mock catalog.
By comparing the results of these simulations with the observational results, they find the best-fit Schechter parameters of \lya\ LFs
\citep[see][for more details of the simulations]{2006PASJ...58..313S, 2008ApJS..176..301O}.
They confirm that the LFs estimated from the simulations are consistent with those derived by the classical method.

\begin{figure}
\centering
\includegraphics[width=8cm]{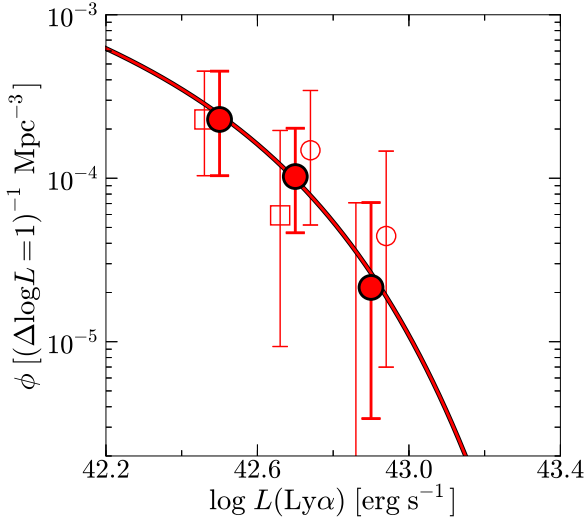}
\caption{\lya\ LF of our $\zi = 7.3$ LAEs.
The red filled circles represent the \lya\ LF derived with the data of the entire fields, i.e. both the SXDS and COSMOS fields.
The red open circles and squares denote our \lya\ LFs estimated with the data of the SXDS and COSMOS fields, respectively.
In the brightest luminosity bin, we also plot the upper error of the \lya\ LF in COSMOS field.
The best-fit Schechter function for the \lya\ LF of the entire fields is shown with the red curve.
} 
\label{fig:LF_thisstudy_compare_z7}
\end{figure}

Figure \ref{fig:LF_thisstudy_compare_z7} presents the \lya\ LF of our $\zi = 7.3$ LAEs in the entire fields
that include both the SXDS and COSMOS fields.
The error bars of this LF include uncertainties from Poisson statistics and cosmic variance.
Here, we estimate the cosmic variance uncertainty, $\sigma_\mathrm{g}$, with
\begin{equation}
\sigma_\mathrm{g} = b_\mathrm{g} \sigma_\mathrm{DM}(\zi , R),
\end{equation}
where $b_\mathrm{g}$ and $\sigma_\mathrm{DM}(\zi , R)$ are the bias parameter and the density fluctuation of dark matter 
at a redshift of $z$ in a radius of $R$, respectively.
We calculate $\sigma_\mathrm{DM}(\zi , R)$ with the growth factor, following \cite{1992ARA&A..30..499C}
with the transfer function given by \cite{1986ApJ...304...15B} \citep[see also][]{2002MNRAS.336..112M}.
Note that the radius of $\sigma_\mathrm{DM}(\zi , R)$ corresponds to that of a sphere which has the survey volume
same as ours, i.e. $2.5 \times 10^{5} \ \mathrm{Mpc}^{3}$.
The value of $\sigma_\mathrm{DM}(\zi , R)$ at $\zi = 7.3$ is estimated to be 0.041. 
Because the bias parameter of $b_\mathrm{g} = 3.6 \pm 0.7$ is obtained for $\zi = 6.6$ LAEs \citep{2010ApJ...723..869O},
we adopt $b_\mathrm{g} \simeq 4$ for $\zi = 7.3$ LAEs under the assumption that $b_\mathrm{g}$ does not significantly evolve at $\zi = 6.6-7.3$.
With this procedure, we estimate the cosmic variance uncertainty to be $\sigma_\mathrm{g}\simeq 0.16$.
In Figure \ref{fig:LF_thisstudy_compare_z7}, we plot the LFs from two independent fields of SXDS and COSMOS.
Because these LFs are consistent within the statistical+cosmic variance uncertainties of the entire-field LF
in two luminosity bins, $\log L_{\mathrm{Ly}\alpha} = 42.7$ and $42.9$ erg s$^{-1}$,
we confirm that our errors of the entire-field LF explain the cosmic variance effects
based on the real observational data of SXDS and COSMOS on the independent sky.

We fit a Shechter function \citep{1976ApJ...203..297S} to our $\zi = 7.3$ \lya\ LF by minimum $\chi^2$ fitting.
The Shechter function is defined by
\begin{equation}
\phi (L) dL = \phi^{*} (L/L^{*})^{\alpha} \exp(-L/L^{*})d(L/L^{*}),
\end{equation}
where 
$\phi^{*}$ and $L^{*}$ represent the characteristic number density and luminosity, respectively,
and $\alpha$ is a power-law slope of the faint-end LF. 
Because the luminosity range of our LF is not wide, the parameter of $\alpha$ in the Schechter function 
cannot be determined.
We fix a power-law slope of $\alpha = -1.5$, which is a fiducial value used for low-$\zi$ \lya\ LFs 
\citep[e.g.,][]{2004ApJ...617L...5M, 2006ApJ...648....7K, 2011ApJ...734..119K, 2008ApJS..176..301O, 2010ApJ...723..869O}.
In the calculations for the $\chi^2$ values, we adopt an upper error as $1\sigma$ in the case that models are beyond the data point of our LF.
Similarly, a lower error is adopted in the case that models are below the data point of our LF.
We obtain the best-fit Schechter parameters of $\phi^{*} = 3.7^{+17.6}_{-3.3} \times 10 ^{-4} \ \mathrm{Mpc}^{3}$ and 
$L^{*}_{\mathrm{Ly}\alpha} = 2.7^{+8.0}_{-1.2} \times 10^{42} \ \mathrm{erg} \ \mathrm{s}^{-1}$ with the fixed $\alpha = -1.5$,
and present these best-fit values in Table \ref{table:schechter}.
The best-fit Schechter function is shown in Figure \ref{fig:LF_thisstudy_compare_z7} with the red solid line.

\subsection{Comparison with $\zi \simeq 7.3$ \lya\ LFs of Previous Studies} \label{sec:compare_z7}

\begin{figure}
\centering
\includegraphics[width=8cm]{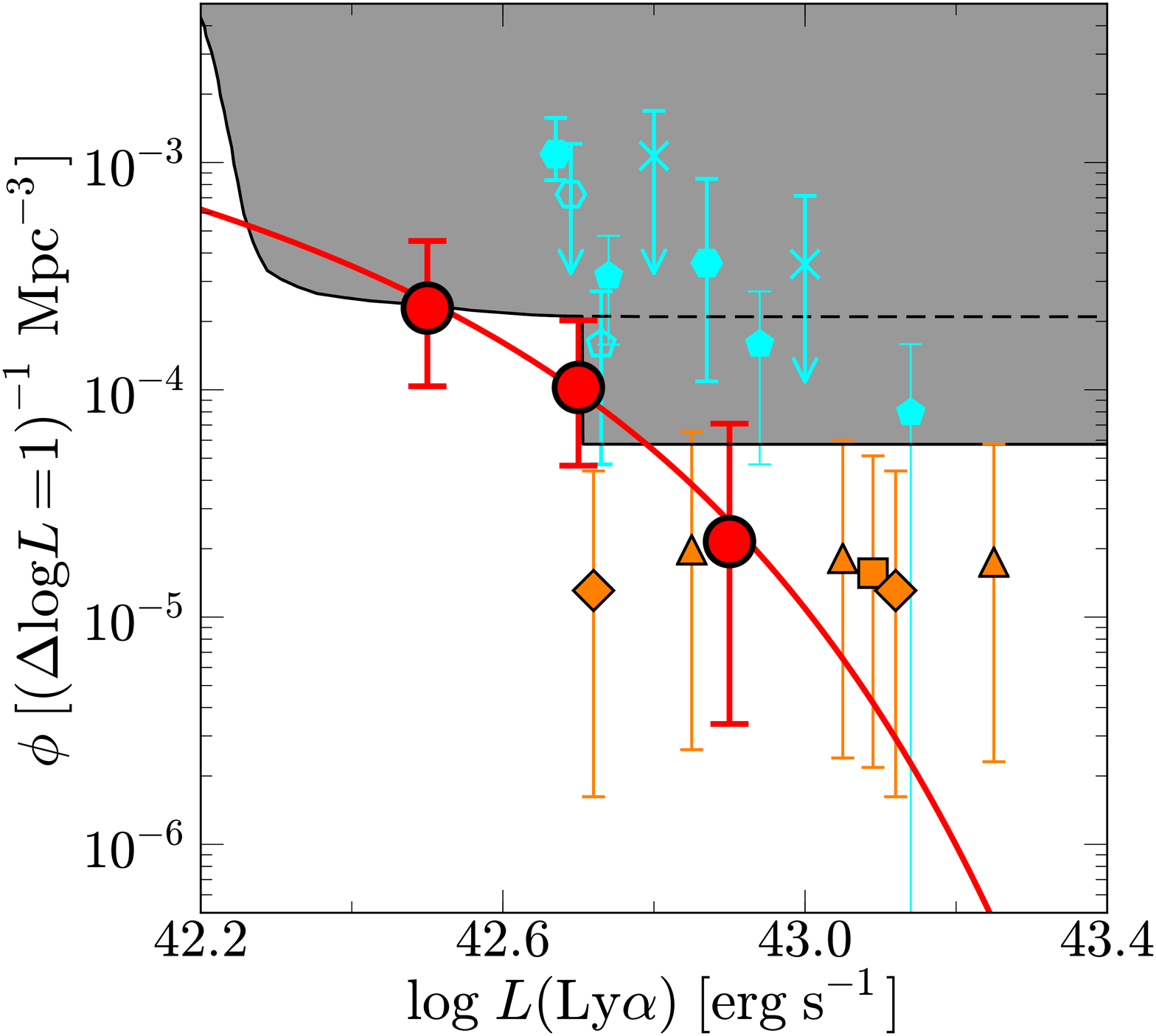}
\caption{Comparison of our $z=7.3$ \lya\ LF with the previous measurements of \lya\ LF at 
$\zi = 7.3\pm 0.4$.
The red circles denote our $z=7.3$ \lya\ LF, and the red curve is the best-fit Schechter function.
The orange diamonds, square, and triangles represent
the Subaru measurements of the \lya\ LF at $z=7.0-7.3$ 
given by \cite{2012ApJ...752..114S}, \cite{2006Natur.443..186I}, and \cite{2010ApJ...722..803O},
respectively.
The gray region indicates the parameter space of $z=7.7$ \lya\ LF 
ruled out by the VLT observations \citep{2012A&A...538A..66C}.
The black dashed line is the upper limit of the number density determined by the VLT photometric observations,
while the black solid line represents the upper limits from the combination of the VLT photometric and spectroscopic data
\citep{2012A&A...538A..66C}.
The cyan filled pentagons, hexagons, and crosses denote 
the 4m-telescope estimates of the \lya\ LF at $z=7.7$
obtained by \cite{2010A&A...515A..97H}, \cite{2012ApJ...745..122K},
and \cite{2010ApJ...721.1853T}, respectively.
The cyan open pentagons and hexagons are the same as
the cyan filled pentagons and hexagons, but for the results of
no emission-line detection of the spectroscopic follow-up observations 
for the 4m-telescope samples,
which are presented in \cite{2012A&A...538A..66C} and \cite{2014ApJ...788...87F},
respectively.
} 
\label{fig:LFcompare_z7p7}
\end{figure}

We compare our $\zi = 7.3$ \lya\ LF with those obtained by previous studies for LAEs at $z=7.0-7.7$,
assuming that the \lya\ LF does not significantly evolve at $\zi=7.3 \pm 0.4$.
In Figure \ref{fig:LFcompare_z7p7},
we plot the previous Subaru measurements of the \lya\ LF at $\zi = 7.0$
\citep{2006Natur.443..186I,2008ApJ...677...12O,2010ApJ...722..803O} and $7.3$ \citep{2012ApJ...752..114S}
that include spectroscopy results.
These previous Subaru results are consistent with
the bright-end of our \lya\ LF within the uncertainties,
while these previous Subaru studies typically reach $L(\mathrm{Ly}\alpha) \sim 10^{43}$ erg s$^{-1}$
that is significantly shallower than our ultra-deep survey.
Similarly, the black solid line of Figure \ref{fig:LFcompare_z7p7} presents the upper limits of the \lya\ LF
given by the VLT observations that identify no LAEs at $z=7.7$ \citep{2012A&A...538A..66C}.
These upper limits of the \lya\ LF are consistent with our results.

On the other hand, we find discrepancies between these Subaru+VLT results including ours and
the previous 4m-telescope results of $z=7.7$ LAEs that are reported by
\citet{2010A&A...515A..97H}, \citet{2010ApJ...721.1853T}, and \citet{2012ApJ...745..122K}.
In Figure \ref{fig:LFcompare_z7p7},
the number densities of the \lya\ LF of the 4m-telescope results are 
about a factor of several or an order of magnitude larger than 
those of the Subaru+VLT results beyond the uncertainties. 
We discuss these discrepancies of $\zi \simeq 7.3$ \lya\ LF measurements
between the Subaru+VLT and 4m-telescope results
in Section \ref{sec:discrepancy}.

\section{discussion} \label{sec:discuss}

\subsection{Discrepancies of $\zi \simeq 7.3$ \lya\ LF Estimates} \label{sec:discrepancy}

In Section \ref{sec:compare_z7}, we find the discrepancies of $z\simeq 7.3$ \lya\ LFs 
between the Subaru+VLT results (including ours) and the 4m-telescope results
\citep{2010A&A...515A..97H,2010ApJ...721.1853T,2012ApJ...745..122K}.
There is a possibility to explain the discrepancies by the cosmic variance effects. 
However, all of these 4m-telescope LF measurements fall above the Subaru+VLT LF estimates.
It is difficult to reconcile all of the 4m-telescope measurements by the chance fluctuations of cosmic variance.
Another possibility is contamination. 
\cite{2012A&A...538A..66C} mention the results of the VLT/X-Shooter spectroscopic follow-up observations for the brightest five 
out of seven photometric LAE candidates of \cite{2010A&A...515A..97H},
and report that no \lya\ emission lines from these \citeauthor{2010A&A...515A..97H}'s candidates
are identified (see J. G. Cuby et al. in preparation).
More recently, \cite{2014ApJ...788...87F} conduct the spectroscopic follow-up observations for the brightest two
out of four photometric candidates of \cite{2012ApJ...745..122K},
and they detect no \lya\ emission line from the \citeauthor{2012ApJ...745..122K}'s candidates 
\citep[see also][]{2013ApJ...771L...6J}.
There is a similar spectroscopic study that reports no detection of \lya\ from $z>7$ LAEs
whose sample is made with 4m-telescope data \citep{2014MNRAS.440.2375M}.
Thus, the photometric samples of \cite{2010A&A...515A..97H} and \cite{2012ApJ...745..122K}
include a significant number of contamination sources that are probably more than a half of
their LAE candidates, which are indicated by the spectroscopic follow-up studies.
Spectroscopic observations for the LAE candidates of \cite{2010ApJ...721.1853T} 
have not been carried out so far. However, it is possible that the \citeauthor{2010ApJ...721.1853T}'s
sample includes a large number of contamination, because of the sample selection from the 4m-telescope data similar to 
those of \citet{2010A&A...515A..97H} and \citet{2012ApJ...745..122K}.
We conclude that our \lya\ LF is consistent with those from the Subaru and VLT studies whose
results are supported by the spectroscopic observations 
\citep{2006Natur.443..186I,2008ApJ...677...12O,2010ApJ...722..803O,2012ApJ...752..114S,2012A&A...538A..66C},
and that our \lya\ LF agrees with the results of the recent deep spectroscopic follow-up observations for
the LAE candidates from the 4m-telescope data \citep{2012A&A...538A..66C,2014ApJ...788...87F,2013ApJ...771L...6J}.

\subsection{Decrease of \lya\ LF from $z=6.6$ to $7.3$} \label{sec:evolution}

\begin{figure}
\centering
\includegraphics[width=8cm]{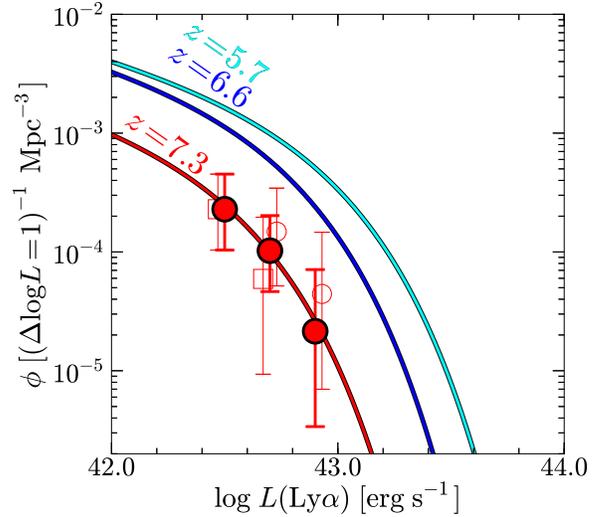}
\caption{Evolution of \lya\ LF at $z=5.7-7.3$. 
The red filled circles are the best estimates of 
our $\zi = 7.3$ \lya\ LF from the data of entire fields.
The red open circles and squares denote our $\zi = 7.3$ \lya\ LFs 
derived with the data of two independent fields of SXDS and COSMOS, respectively.
The red curve is the best-fit Schechter function for the best estimate of our $z=7.3$ \lya\ LF.
The cyan and blue curves are the best-fit Schechter functions of the \lya\ LFs
at $z=5.7$ and $6.6$ obtained by \cite{2008ApJS..176..301O} and \cite{2010ApJ...723..869O}, respectively.
} 
\label{fig:LFcompare_lowz}
\end{figure}

In this section, we examine whether the \lya\ LF evolves from $\zi = 6.6$ to $7.3$.
As described in Section \ref{sec:z7p3_cand},
we reach the \lya\ limiting luminosity of $2.4 \times 10^{42}$ erg s$^{-1}$
that is comparable with those of previous Subaru $\zi = 3.1-6.6$ studies
\citep{2006PASJ...58..313S, 2006ApJ...648....7K, 2011ApJ...734..119K, 2008ApJS..176..301O, 2010ApJ...723..869O, 2010ApJ...725..394H}.
Moreover, the size of survey area, $\simeq 0.5$ deg$^2$, is comparable with these Subaru studies.
Our ultra-deep observations in the large areas allow us to perform a fair comparison of the \lya\ LFs at different redshifts.
We compare our \lya\ LF at $\zi = 7.3$ with those at $\zi = 5.7$ and $6.6$ in Figure \ref{fig:LFcompare_lowz},
and summarize the best-fit Schechter parameters at $\zi =5.7$, $6.6$, and $7.3$ in Table \ref{table:schechter}.
For the $z=5.7$ and $6.6$ data, we use the \lya\ LF measurements of \citet{2010ApJ...723..869O} 
derived from the largest LAE samples, to date, at these redshifts, and the \lya\ LF measurements include
all of the major Subaru survey data \citep{2006PASJ...58..313S, 2006ApJ...648....7K, 2011ApJ...734..119K}
and the cosmic variance uncertainties in their errors. Nevertheless, the difference of the best-estimate \lya\ LFs
is negligibly small between these studies.
In Figure \ref{fig:LFcompare_lowz}, we find
a significant decrease of the \lya\ LFs from $\zi =6.6$ to $7.3$ largely beyond the error bars.
In our survey, we expect to find $65$ $z=7.3$ LAEs in the case of no \lya\ LF evolution from $\zi = 6.6$ to $7.3$,
but identify only 7 $z=7.3$ LAEs by our observations
that are about an order of magnitude smaller than the expected LAEs.

\begin{figure}
\centering
\includegraphics[width=8cm]{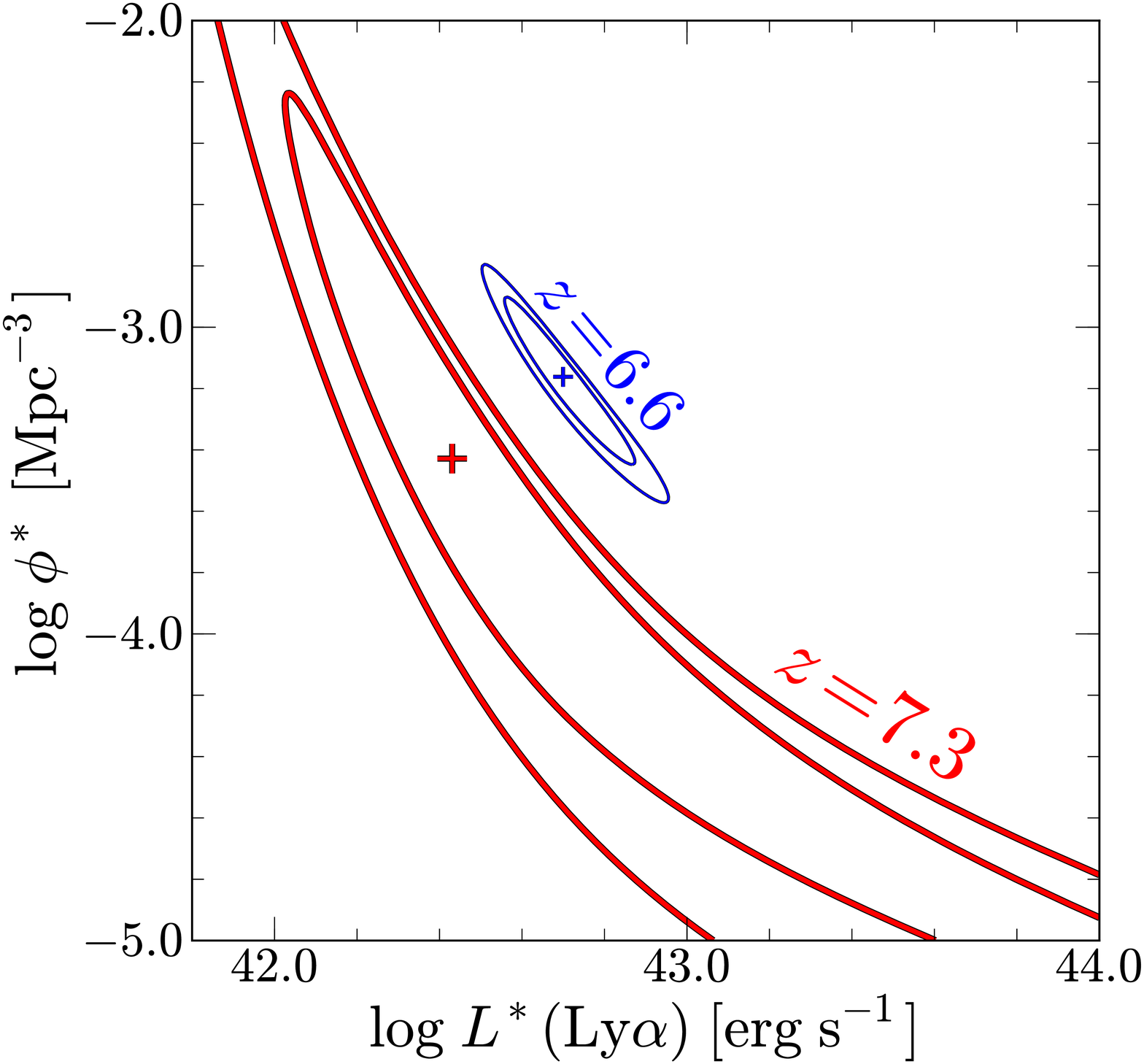}
\caption{Error contours of Schechter parameters, $L^{*}_{\mathrm{Ly}\alpha}$ and $\phi^{*}$.
The red contours represent our \lya\ LF at $z=7.3$, while the blue contours denote
the one at $z=6.6$ obtained by \cite{2010ApJ...723..869O}.
The inner and outer contours indicate the $68 \%$ and $90 \%$ confidence levels, respectively.
The red and blue crosses show the best-fit Schechter parameters for the \lya\ LFs at $\zi = 7.3$ and $6.6$, respectively.
} 
\label{fig:error_cont}
\end{figure}

To quantify this evolution, we evaluate the error distribution of Schechter parameters. 
Because we fix the Schechter parameter of $\alpha$ to $-1.5$,
we examine the error distribution of $L^{*}_{\mathrm{Ly}\alpha}$ and $\phi^{*}$ with the fixed value of $\alpha=-1.5$.
Figure \ref{fig:error_cont} shows error contours of the Schechter parameters of our $\zi = 7.3$ \lya\ LF, together with those of $\zi = 6.6$ LF
\citep{2010ApJ...723..869O}.
Our measurements indicate that 
the Schechter parameters of $\zi = 7.3$ LF are different from those of $\zi = 6.6$ \lya\ LF, 
and that the \lya\ LF decreases from $\zi = 6.6$ to $7.3$ at the $> 90 \%$ confidence level.
Because our $\zi = 7.3$ \lya\ LF is derived with the same procedures
as the $z=6.6$ \lya\ LF \citep{2010ApJ...723..869O},
one expects no systematic errors raised by the analysis technique
for the comparison of the $z=6.6$ and $7.3$ results.
From this aspect, it is reliable that the \lya\ LF declines from $\zi =6.6$ to $7.3$ significantly.
Here, we also discuss the possibilities of the LF decrease mimicked by our sample biases.
In section \ref{sec:contami}, we assume that there is no contamination in our $\zi =7.3$ LAE sample.
If there exist some contamination sources, the $z=7.3$ \lya\ LF corrected for contamination
should fall below the present estimate of the $z=7.3$ \lya\ LF. In this case, 
our conclusion of the significant LF decrease is even strengthened.
In Section \ref{sec:z7p3_cand}, we define the selection criterion of the rest-frame \lya\ equivalent width of 
EW$_{0} \gtrsim 0 \mathrm{\AA}$ for our $\zi = 7.3$ LAEs. This criterion of the EW$_0$ limit is slightly different
from that of the LAEs for the $\zi =6.6$ \lya\ LF estimates. However, the EW$_0$ limit for the
$z=6.6$ LAEs is EW$_0\gtrsim 14$\AA\ \citep{2010ApJ...723..869O} that is larger than 
our EW$_0$ limit of $\zi = 7.3$ LAEs. Because our EW$_0$ limit gives more $z=7.3$ LAEs
to our sample than that of $z=6.6$ LAEs, 
the conclusion of the \lya\ LF decrease from $z=6.6$ to $7.3$ is unchanged
by the EW$_0$ limit.

\begin{deluxetable*}{cccccc}
\tabletypesize{\footnotesize}
\tablecaption{Best-fit Schechter Parameters and \lya\ Luminosity Densities 	\label{table:schechter}}
\tablewidth{0pt}
\tablehead{
\multicolumn{1}{c}{} & 
\multicolumn{1}{c}{$L^{*}_{\mathrm{Ly}\alpha}$}	&
\multicolumn{1}{c}{$\phi^{*}$}	&
\multicolumn{1}{c}{$\rho^{\mathrm{Ly}\alpha}$ \tablenotemark{a}}	&
\multicolumn{1}{c}{$\rho^{\mathrm{Ly}\alpha,\mathrm{tot}}$ \tablenotemark{b}}	&
\multicolumn{1}{c}{}	\\
\multicolumn{1}{c}{Redshift} & 
\multicolumn{1}{c}{($10^{42}$ erg $\mathrm{s}^{-1}$)}	&
\multicolumn{1}{c}{($10^{-4}$ $\mathrm{Mpc}^{-3}$)}	&
\multicolumn{1}{c}{($10^{39}$ erg s$^{-1}$ $\mathrm{Mpc}^{-3}$)}	&
\multicolumn{1}{c}{($10^{39}$ erg s$^{-1}$ $\mathrm{Mpc}^{-3}$)}	&
\multicolumn{1}{c}{Reference}
}
\startdata
5.7	& $6.8^{+3.0}_{-2.1}$	& $7.7^{+7.4}_{-3.9}$	& $3.6^{+3.1}_{-1.7}$	& $9.2^{+6.6}_{-3.7}$	& \cite{2008ApJS..176..301O}	\\
6.6	& $4.4^{+0.6}_{-0.6}$	& $8.5^{+3.0}_{-2.2}$	& $1.9^{+0.5}_{-0.4}$	& $6.6^{+1.0}_{-0.8}$	& \cite{2010ApJ...723..869O}	\\
7.3	& $2.7^{+8.0}_{-1.2}$	& $3.7^{+17.6}_{-3.3}$	& $0.31^{+0.19}_{-0.12}$	& $1.8^{+3.8}_{-1.1}$	& This study
\enddata
\tablenotetext{a}{\lya\ luminosity densities integrated down to the observation luminosity limit, $\log L_{\mathrm{Ly}\alpha}=42.4$ erg s$^{-1}$, for all of the redshifts.}
\tablenotetext{b}{Total \lya\ luminosity densities integrated down to $L_{\mathrm{Ly}\alpha}=0$.}
\end{deluxetable*}

\subsection{Accelerated Evolution of \lya\ LF at $\zi \gtrsim 7$} \label{sec:acc_evolution}

\begin{figure*}
\centering
\includegraphics[width=0.70\textwidth]{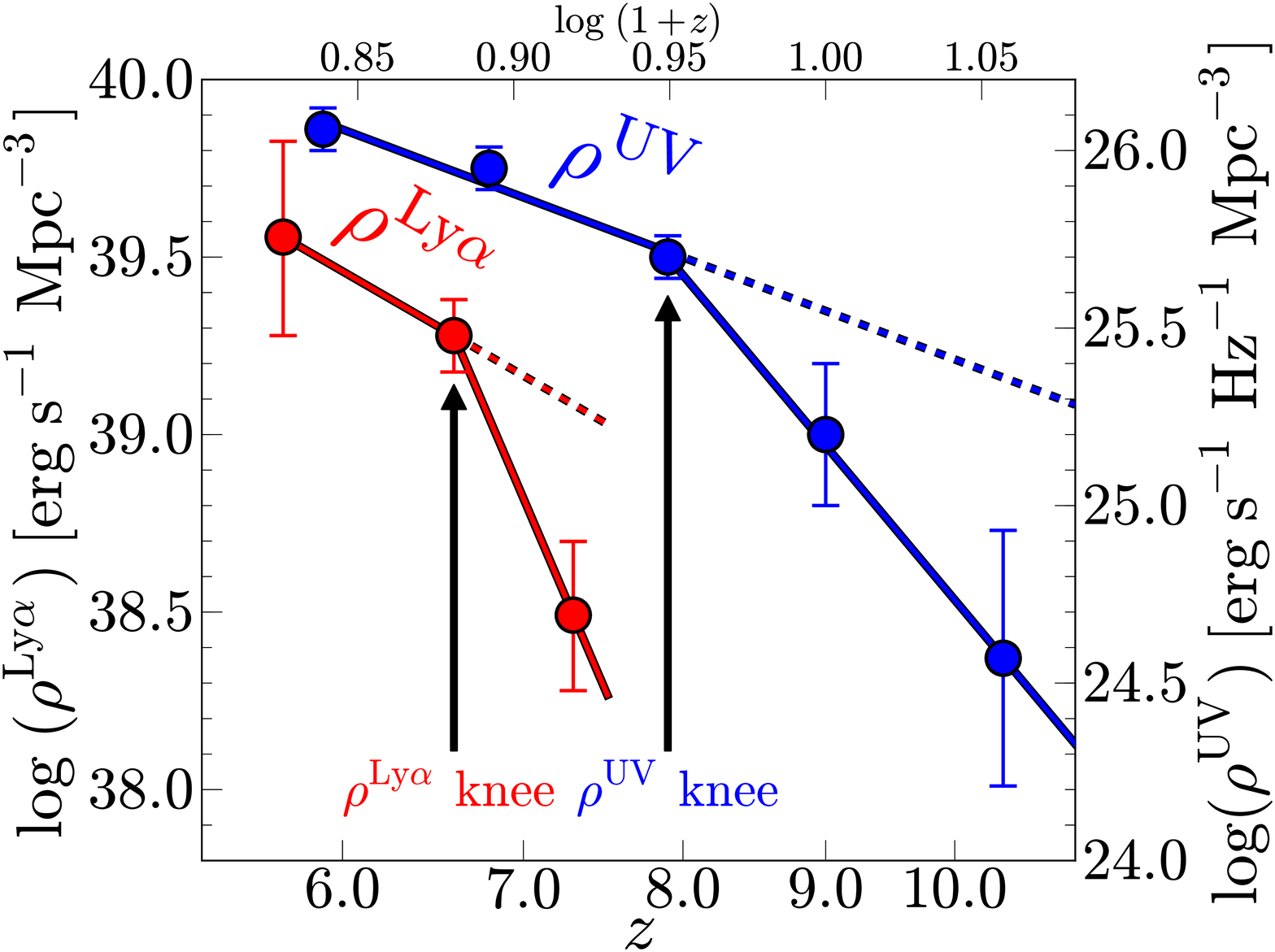}
\caption{Evolution of \lya\ and UV luminosity densities.
The red circles are the \lya\ luminosity densities
obtained by this study, \cite{2010ApJ...723..869O}, and \cite{2008ApJS..176..301O}
for $z=7.3$, $6.6$, and $5.7$, respectively.
The blue circles are the UV luminosity densities
given by \cite{2014arXiv1403.4295B} for $z=5.9$, $6.8$, $7.9$, and $10.4$,
and \cite{2013ApJ...763L...7E} for $z=9.0$.
The left ordinate axis is referred for the \lya\ luminosity densities,
and the right ordinate axis is for the UV luminosity densities.
The \lya\ luminosity density starts evolving acceleratingly at $z\sim 7$,
while the UV luminosity density rapidly decreases at $z\sim 8$ and beyond.
The $\rho^{{\rm Ly}\alpha}$ and $\rho^{\rm UV}$ knees are indicated with the arrows.
}
\label{fig:rho_evo}
\end{figure*}

Figure \ref{fig:LFcompare_lowz} implies that the decrease of the \lya\ LF from $\zi = 6.6$ to $7.3$
is larger than that from $\zi = 5.7$ to $6.6$, i.e., there is an accelerated evolution of the \lya\ LF at $\zi = 6.6 -7.3$.
To evaluate this evolution quantitatively, we calculate the \lya\ luminosity densities, $\rho^{\mathrm{Ly}\alpha}$,
down to the common luminosity limit of $\log L_{\rm Ly\alpha}=42.4$ erg s$^{-1}$ reached by the observations
for LAEs at $\zi = 5.7$, $6.6$, and $7.3$. Similarly, we estimate the total \lya\ luminosity densities, 
$\rho^{\mathrm{Ly}\alpha, \mathrm{tot}}$, that are integrated down to $L_{\rm Ly\alpha}=0$
with the best-fit Schechter functions.
Figure \ref{fig:rho_evo} presents the evolution of $\rho^{\mathrm{Ly}\alpha}$,
and Table \ref{table:schechter} summarizes the values of these \lya\ luminosity densities
at each redshift.
Here, we use $\log (1+\zi)$ for the abscissa in Figure \ref{fig:rho_evo},
because we compare the evolution of $\rho^{\mathrm{Ly}\alpha}$
with that of UV luminosity densities, $\rho^{\mathrm{UV}}$, derived by \cite{2013ApJ...773...75O} who use $\log (1+\zi)$
(see Section \ref{sec:imply}).
In this figure, we find a rapid decrease of the \lya\ luminosity density at $\zi = 6.6-7.3$.
To quantify this evolution, we calculate ratios of $\rho^{\mathrm{Ly}\alpha}_{\zi_2} / \rho^{\mathrm{Ly}\alpha}_{\zi_1}$
that are shown in Table \ref{table:rho_evo}, where $\zi_1$ and $\zi_2$ are redshifts.
We fit the evolution of $\rho^{\mathrm{Ly}\alpha}_{\zi}$ to the power-law function,
\begin{equation}
	\rho^{\mathrm{Ly}\alpha}_{\zi} \propto (1+\zi)^{n(\rho)},
\label{eq:rho_evolution}
\end{equation}
and obtain $n(\rho) = -5.0^{+4.2}_{-9.5}$ at $\zi = 5.7 - 6.6$ and $n(\rho) = -20.8^{+5.1}_{-9.4}$ at $\zi = 6.6 - 7.3$.
Because these values of $n(\rho)$ are significantly different,
we conclude that the \lya\ LF evolves acceleratingly at $\zi \gtrsim 7$.

\begin{deluxetable}{ccc}
\tabletypesize{\footnotesize}
\tablecaption{\lya\ Luminosity Density Evolution \label{table:rho_evo}}
\tablewidth{0pt}
\tablehead{
\colhead{Redshift Range} & 
\colhead{$\rho^{\mathrm{Ly}\alpha}$ Evolution\tablenotemark{a}} &
\colhead{} \\
\colhead{$z = \zi_{1} - \zi_2$} & 
\colhead{$\rho^{\mathrm{Ly}\alpha}_{\zi_2} / \rho^{\mathrm{Ly}\alpha}_{\zi_1}$} &
\colhead{$n (\rho)$\tablenotemark{b}} 
}
\startdata
$\zi = 5.7 - 6.6$	& $0.53 \pm 0.37$	& $-5.0^{+4.2}_{-9.5}$	\\
$\zi = 6.6 - 7.3$	& $0.16 \pm 0.09$	& $-20.8^{+5.1}_{-9.4}$	\\
\addlinespace[2pt]
\hline
\addlinespace[2pt]
$\zi = 5.7 - 7.3$	& $0.09 \pm 0.07$	& $-11.2^{+2.7}_{-7.0}$	
\enddata
\tablenotetext{a}{Best-fit values of luminosity density ratio, $\rho^{\mathrm{Ly}\alpha}_{\zi_2}/\rho^{\mathrm{Ly}\alpha}_{\zi_1}$, where the indices of $\zi_1$ and $\zi_2$ denote redshifts.}
\tablenotetext{b}{Power-law slope $n(\rho)$ defined with Equation (\ref{eq:rho_evolution}).}
\end{deluxetable}

\begin{figure*}
\centering
\includegraphics[width=1.0\textwidth]{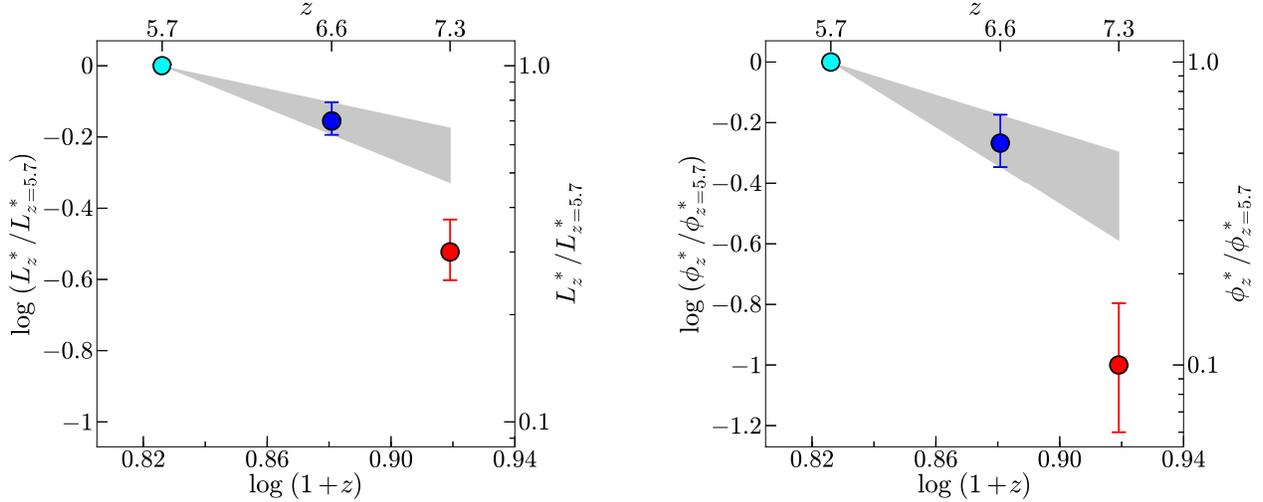}
\caption{
$L^{*}_{\mathrm{Ly}\alpha}$ and $\phi^{*}$ as a function of redshift
for the cases of pure-luminosity evolution (left panel) and pure-number density evolution (right panel).
For the both cases, the values of $z=5.7$ are used for the normalization.
The gray shades denote the $L^{*}$ or $\phi^{*}$ evolution at $\zi= 5.7-6.6$ with the uncertainties,
and these evolutionary trends are extrapolated to $\zi =7.3$.
} 
\label{fig:pure_evo}
\end{figure*}

We also investigate pure-luminosity and number density evolution cases
to test whether this rapid \lya\ LF evolution is dominated by a $L^{*}$ or $\phi^{*}$ decrease.
These evolution cases are examined by the minimum $\chi^2$ fitting.
For example, to evaluate the pure-luminosity evolution from $\zi = 6.6$ to $7.3$, 
we take a set of three parameters of $L^{*}_{\zi = 6.6}$, $L^{*}_{\zi = 7.3}/L^{*}_{\zi = 6.6}$, and $\phi^{*}$,
where $L^{*}_{\zi = 6.6}$ and $L^{*}_{\zi = 7.3}/L^{*}_{\zi = 6.6}$ are a Schechter parameter of $L^{*}$ at $\zi = 6.6$
and a ratio of $z=7.3$ $L^{*}$ to $z=6.6$ $L^*$, respectively.
Here, $\phi^{*}$ is a common value in $\zi = 6.6$ and $7.3$, and the Schechter parameter of $\alpha$ is fixed to $-1.5$.
We prepare Schechter functions at $\zi = 6.6$ and $7.3$ with the sets of three parameters,
and search for the best-fit parameters that minimize $\chi^{2}$ by the simultaneous fit 
of Schechter functions to $\zi =6.6$ and $7.3$ \lya\ LFs.
In this way, we obtain the best-fit parameter of $L^{*}_{\zi = 7.3}/L^{*}_{\zi = 6.6}$ 
that corresponds to a fraction of $L^*$ for the pure-luminosity evolution between $\zi = 6.6$ and $7.3$.
Similarly, we estimate $L^{*}_{\zi = 7.3}/L^{*}_{\zi = 5.7}$ and $L^{*}_{\zi = 6.6}/L^{*}_{\zi = 5.7}$
at the redshift ranges. We also evaluate the pure-number density evolution with the ratios of $\phi^*$ 
in the same manner.
We summarize the best-fit parameters for these pure-luminosity and number density evolutions
in Table \ref{table:Lphi_evo}.
Figure \ref{fig:pure_evo} shows the evolutions of the $L^*$ and $\phi^*$ ratios
from $z=5.7$ to a redshift of $z$.
In Figure \ref{fig:pure_evo}, the shaded area
denotes the $L^{*}$ (and $\phi^{*}$) evolution at $\zi = 5.7 - 6.6$ with the measurement uncertainties,
and indicates the extrapolation of this evolutionary trend to $z=7.3$.
We find that the ratios of $L^{*}$ and $\phi^{*}$ drop from $z=6.6$ to $7.3$ 
below the shaded area. Similar to Equation (\ref{eq:rho_evolution}),
we approximate the pure $L^{*}$ and $\phi^{*}$ evolutions by power laws
whose indices are $n(L^*)$ and $n(\phi^*)$:
\begin{equation}
	\begin{split}
		L^{*}_{\zi} \propto (1 + \zi )^{n(L^{*})}	\\
		\phi^{*}_{\zi} \propto (1 + \zi )^{n(\phi^{*})}.	\label{eq:pure_evolution}
	\end{split}
\end{equation}
We summarize the best-fit $n(L^{*})$ and $n(\phi^{*})$ values in Table \ref{table:Lphi_evo}.
In either case of the pure-luminosity or number density evolution, the indices of $n$ 
at $z=6.6-7.3$ is significantly smaller than those at $z=5.7-6.6$.
These results are consistent with our conclusion of the accelerated evolution of the \lya\ LF at $z\gtrsim 7$.
The $\chi^{2}$ values are comparable for the pure-luminosity and number density evolution
cases (see Table \ref{table:Lphi_evo}), although
the $\chi^{2}$ values of the pure-luminosity evolution are
slightly smaller than those of the pure-number density evolution.
The available \lya\ LF data do not have accuracies to discuss 
the dominant component of the evolution at $z=5.7-7.3$.
Nevertheless, if we assume the \lya\ LF evolution is dominated by a pure $L^*$ evolution
whose $\chi^2$ values are smaller than those of a pure $\phi^*$ evolution, 
we find that, in the pure $L^*$ evolution,
the decreases of the \lya\ LF are $30\% \ (=[1 - L^{*}_{\zi = 6.6} / L^{*}_{\zi = 5.7}] \times 100)$ 
and $70\% \ (= [1 - L^{*}_{\zi = 7.3} / L^{*}_{\zi = 5.7}] \times 100)$ at $\zi = 5.7 - 6.6$ and $\zi = 5.7 - 7.3$, respectively.
In other words, the typical LAE has gotten brighter by 1.4 times from $\zi = 6.6$ to $5.7$
and 3.3 times from $\zi = 7.3$ to $5.7$.

\begin{deluxetable*}{ccccccc}
\tabletypesize{\footnotesize}
\tablecaption{Best-Fit Parameters for Pure-Luminosity and Number Density evolution Cases \label{table:Lphi_evo}}
\tablewidth{0pt}
\tablehead{
\colhead{Redshift Range} & 
\colhead{$L^{*}_{\mathrm{Ly}\alpha}$ Evolution\tablenotemark{a}} &
\colhead{} &
\colhead{} &
\colhead{$\phi^{*}$ Evolution\tablenotemark{d}} &
\colhead{}&
\colhead{}	\\
\colhead{$\zi_{1} - \zi_2$} & 
\colhead{$L^{*}_{\zi_2} / L^{*}_{\zi_1}$} &
\colhead{$\chi^{2} (L^{*})$\tablenotemark{b}} &
\colhead{$n (L^{*})$\tablenotemark{c}} &
\colhead{$\phi^{*}_{\zi_2} / \phi^{*}_{\zi_1}$} &
\colhead{$\chi^{2} (\phi^{*})$\tablenotemark{e}}&
\colhead{$n (\phi^{*})$\tablenotemark{f}}
}
\startdata
$\zi = 5.7 - 6.6$	& $0.70^{+0.09}_{-0.06}$	& $4.2$	& $-2.8^{+1.0}_{-0.7}$	& $0.54^{+0.13}_{-0.09}$	& $4.7$	& $-4.9^{+1.7}_{-1.4}$	\\
$\zi = 6.6 - 7.3$	& $0.42^{+0.09}_{-0.07}$	& $1.6$	& $-9.8^{+2.2}_{-2.0}$	& $0.18^{+0.09}_{-0.07}$	& $1.8$	& $-19.5^{+4.6}_{-5.6}$	\\
\addlinespace[2pt]
\hline
\addlinespace[2pt]
$\zi = 5.7 - 7.3$	& $0.30^{+0.07}_{-0.05}$	& $2.9$	& $-5.6^{+1.0}_{-0.9}$	& $0.10^{+0.06}_{-0.04}$	& $3.4$	& $-10.8^{+2.2}_{-2.4}$
\enddata
\tablenotetext{a}{Best-fit value of $L^{*}_{\zi_2}/L^{*}_{\zi_1}$, where the indices of $\zi_1$ and $\zi_2$ indicate redshifts.}
\tablenotetext{b}{$\chi^{2}$ for the best-fit $L^{*}_{\zi_2}/L^{*}_{\zi_1}$.}
\tablenotetext{c}{Power-law slope $n(L^*)$ of Equation (\ref{eq:pure_evolution}) for pure-luminosity evolution case.}
\tablenotetext{d}{Best-fit value of $\phi^{*}_{\zi_2}/\phi^{*}_{\zi_1}$.}
\tablenotetext{e}{$\chi^{2}$ for the best-fit $\phi^{*}_{\zi_2}/\phi^{*}_{\zi_1}$.}
\tablenotetext{f}{Power-law slope $n(\phi^*)$ of Equation (\ref{eq:pure_evolution}) for pure-number density evolution case.}
\end{deluxetable*}

\subsection{Implications from the Accelerated Evolution of \lya\ LF} \label{sec:imply}

In Section \ref{sec:acc_evolution}, we find that the \lya\ LF shows the accelerated evolution at a redshift beyond $\zi \sim 7$.
We refer to the redshift starting the rapid decrease of the \lya\ luminosity density as ``$\rho^{\rm Ly\alpha}$ knee''
that is indicated in Figure \ref{fig:rho_evo}. In contrast with this evolution of the \lya\ LF,
there is no such a rapid decrease in the UV LF at $z\sim 7$, but only at $z>8$, 
if any (\citealt{2013ApJ...773...75O, 2014arXiv1403.4295B}).
Figure \ref{fig:rho_evo} compares the evolution of $\rho^{\rm Ly\alpha}$ (red symbols)
and $\rho^{\rm UV}$ (blue symbols). Although the rapid decrease of
the UV LF at $z>8$ is still an open question (see, e.g., \citealt{2013ApJ...763L...7E,2013ApJ...768...71R}), 
we refer to the redshift starting this possible rapid decrease of the UV LF as ``$\rho^{\rm UV}$ knee''.
Again, there is a significant redshift difference between $\rho^{\rm Ly\alpha}$ and $\rho^{\rm UV}$ knees 
(Figure \ref{fig:rho_evo}).
Because the evolution of $\rho^{\rm UV}$ correlates with the cosmic star-formation rate (SFR) history,
the accelerated evolution of the \lya\ LF found at $z\sim 7$ is not originated from a rapid decrease of the SFR density.
To explain this accelerated evolution of the \lya\ LF, there should exist physical mechanisms related to the \lya\ production and 
escape processes.
The simple interpretation of the \lya\ LF decrease is that the \lya\ damping wing of IGM given by cosmic reionization 
absorbs \lya\ of galaxies strongly towards high redshifts.
Here, we first investigate this simple scenario of cosmic reionization in Sections \ref{sec:xHI}-\ref{sec:compare_tau},
and then discuss the physical origin of the accelerated evolution of the \lya\ LF with
various possible scenarios in Section \ref{sec:physical_origin}.

\subsubsection{Constraints on \xhi\ at $\zi = 7.3$} \label{sec:xHI}

In Sections \ref{sec:xHI}-\ref{sec:compare_tau}, 
we discuss the simple scenario of cosmic reionization that contributes 
to the accelerated evolution of the \lya\ LF.
We define $T^{\mathrm{IGM}}_{\mathrm{Ly}\alpha, \zi}$ as a \lya\ transmission through IGM at a redshift of $z$,
and calculate $T^{\mathrm{IGM}}_{\mathrm{Ly}\alpha, \zi = 7.3} / T^{\mathrm{IGM}}_{\mathrm{Ly}\alpha, \zi = 5.7}$ 
to estimate \xhi\ at $\zi = 7.3$. Because cosmic reionization has been completed at $\zi = 5.7$ \citep{2006AJ....132..117F},
the \lya\ damping wing absorption of IGM is negligible at $z=5.7$.

Section \ref{sec:acc_evolution} presents the estimates of 
the \lya\ luminosity densities from the \lya\ LFs of \cite{2008ApJS..176..301O} and this study
at $\zi = 5.7$ and $7.3$, respectively (Table \ref{table:schechter}).  
There are two estimates of the \lya\ luminosity densities,
the observed \lya\ luminosity density, $\rho^{\mathrm{Ly}\alpha}$,
and the total one, $\rho^{\mathrm{Ly}\alpha, \mathrm{tot}}$.
\cite{2010ApJ...723..869O} calculate these two \lya\ luminosity densities with their data and confirm that 
the ratio of these different estimates agree within the error bars (see Figure 19 of \citealt{2010ApJ...723..869O}).
Thus, we adopt the total \lya\ luminosity density for our fiducial results for cosmic reionization,
which include no systematic bias from observations.
With the values shown in Table \ref{table:schechter},
we obtain $\rho^{\mathrm{Ly}\alpha, \mathrm{tot}}_{\zi = 7.3} / \rho^{\mathrm{Ly}\alpha, \mathrm{tot}}_{\zi = 5.7} = 0.20$.

Because the \lya\ LF evolution is made not only by cosmic reionization, but also by the SFR change of galaxy evolution,
we subtract the effect of the SFR density evolution from the \lya\ luminosity density evolution.
An SFR of galaxy is correlated with the UV luminosity.
The UV luminosity of $\zi = 7.3$ LAE in principle can be estimated by the subtraction of the \lya\ line flux 
from the $\textit{z}'$-band flux.
However, we cannot derive the reliable UV luminosities from our data.
This is because there exist the large uncertainties of the $\textit{z}'$-band magnitude and 
the contamination of unknown amount of IGM absorption
that make a significant bias in the estimate of the UV continuum
as demonstrated in the simulations of \citet{2006PASJ...58..313S}.
To derive a reliable UV LF of $\zi \gtrsim 6$ LAEs, one needs deep near-infrared data,
such as $\textit{J}$ and $\textit{H}$ images,
which cover the continuum emission longward of the \lya\ line for most of LAEs, 
but no such data are available for LAE studies, to date.
Since we cannot derive 
a reliable UV LF of $\zi = 7.3$ LAE from our data,
we quantify $\rho^{\mathrm{UV}}$
at $\zi = 5.7 - 7.3$ given by the other observations.
We use $\rho_{\mathrm{UV}}$ measured with
the samples of dropout galaxies \citep{2009ApJ...705..936B, 2011ApJ...737...90B},
and estimate $\rho^{\mathrm{UV}}$ at $\zi = 5.7$ and $7.3$ by the interpolation of this evolution.
We, thus, obtain $\rho^{\mathrm{UV}}_{\zi = 7.3} / \rho^{\mathrm{UV}}_{\zi = 5.7} = 0.70$.

Following the procedure of \citet{2010ApJ...723..869O},
we estimate $T^{\mathrm{IGM}}_{\mathrm{Ly}\alpha, \zi = 7.3} / T^{\mathrm{IGM}}_{\mathrm{Ly}\alpha, \zi = 5.7}$ and \xhi.
The value of $\rho^{\mathrm{Ly}\alpha}$ is given by
\begin{equation}
\rho^{\mathrm{Ly}\alpha} = \kappa \ T^{\mathrm{IGM}}_{\mathrm{Ly}\alpha} \ f^{\mathrm{esc}}_{\mathrm{Ly}\alpha} \ \rho^{\mathrm{UV}}, \label{eq:eq1}
\end{equation}
where $\kappa$ is a factor converting from UV to \lya\ luminosities, which depends on stellar population.
$f^{\mathrm{esc}}_{\mathrm{Ly}\alpha}$ is a fraction of \lya\ emission escape from a galaxy through the inter-stellar medium (ISM)
absorption including galactic neutral hydrogen and dust attenuation.
With Equation (\ref{eq:eq1}), $T^{\mathrm{IGM}}_{\mathrm{Ly}\alpha, \zi = 7.3} / T^{\mathrm{IGM}}_{\mathrm{Ly}\alpha, \zi = 5.7}$ 
is written as
\begin{equation}
\frac{T^{\mathrm{IGM}}_{\mathrm{Ly}\alpha, \zi = 7.3}}{T^{\mathrm{IGM}}_{\mathrm{Ly}\alpha, \zi = 5.7}} = \frac{\kappa_{\zi = 5.7}}{\kappa_{\zi = 7.3}} \frac{f^{\mathrm{esc}}_{\mathrm{Ly}\alpha, \zi = 5.7}}{f^{\mathrm{esc}}_{\mathrm{Ly}\alpha, \zi = 7.3}} \frac{\rho^{\mathrm{Ly}\alpha, \mathrm{tot}}_{\zi = 7.3} / \rho^{\mathrm{Ly}\alpha, \mathrm{tot}}_{\zi = 5.7}}{\rho^{\mathrm{UV}}_{\zi = 7.3} / \rho^{\mathrm{UV}}_{\zi = 5.7}}.	\label{eq:eq2}
\end{equation}
Assuming that the stellar population of LAEs is the same at $\zi = 5.7$ and $7.3$ (i.e., $\kappa_{\zi = 5.7}/\kappa_{\zi = 7.3} = 1$),
and the physical state of ISM is not evolved at $\zi = 5.7 - 7.3$
(i.e., $f^{\mathrm{esc}}_{\mathrm{Ly}\alpha, \zi = 5.7} / f^{\mathrm{esc}}_{\mathrm{Ly}\alpha, \zi = 7.3} = 1$),
we obtain
\begin{equation}
\frac{T^{\mathrm{IGM}}_{\mathrm{Ly}\alpha, \zi = 7.3}}{T^{\mathrm{IGM}}_{\mathrm{Ly}\alpha, \zi = 5.7}} = \frac{\rho^{\mathrm{Ly}\alpha, \mathrm{tot}}_{\zi = 7.3} / \rho^{\mathrm{Ly}\alpha, \mathrm{tot}}_{\zi = 5.7}}{\rho^{\mathrm{UV}}_{\zi = 7.3} / \rho^{\mathrm{UV}}_{\zi = 5.7}}.
\end{equation}
From the ratios of the \lya\ and UV luminosity densities described above, 
we estimate $T^{\mathrm{IGM}}_{\mathrm{Ly}\alpha, \zi = 7.3} / T^{\mathrm{IGM}}_{\mathrm{Ly}\alpha, \zi = 5.7} $ to be $0.29$.

We use theoretical models to constrain \xhi\ at $\zi = 7.3$ with our estimates of 
$T^{\mathrm{IGM}}_{\mathrm{Ly}\alpha, \zi = 7.3} / T^{\mathrm{IGM}}_{\mathrm{Ly}\alpha, \zi = 5.7} = 0.29$.
In the analytic model of \cite{2004MNRAS.349.1137S}, the \lya\ transmission fraction of IGM is related to \xhi\
in two cases of no galactic wind and a galactic outflow that give shifts of \lya\ line from a systemic velocity 
by $0$ and $360\ \mathrm{km}/\mathrm{s}$, respectively.
The value of $T^{\mathrm{IGM}}_{\mathrm{Ly}\alpha, \zi = 7.3} / T^{\mathrm{IGM}}_{\mathrm{Ly}\alpha, \zi = 5.7} = 0.29$
corresponds to \xhi\ $\sim 0.0$ and $\sim 0.8$ in the former and the latter case, respectively.
Because recent studies have reported that the \lya\ line emission of LAE at $\zi = 2.2 $ is redshifted 
by $\sim 200$ km/s \citep{2013ApJ...765...70H, 2014ApJ...788...74S}, we take \xhi\ $\sim 0.5$
that is the \xhi\ value interpolated by the \lya\ velocity shift
in Figure 25 of \cite{2004MNRAS.349.1137S}.
\cite{2007MNRAS.381...75M} predict \lya\ LFs for various \xhi\ values with radiative transfer simulations.
By the comparison of our \lya\ LF with these simulation results
in Figure 4 of \cite{2007MNRAS.381...75M}, 
we obtain \xhi\ $\sim 0.7$.
In the models of \cite{2007MNRAS.377.1175D, 2007MNRAS.379..253D}, the \lya\ transmission fraction of IGM is related to the size of typical ionized bubbles,
and \cite{2006MNRAS.365.1012F} predict \xhi\ from the size of the ionized bubble with the analytic model.
Based on Figure 6 of \cite{2007MNRAS.379..253D},
our estimates of the \lya\ transmission fraction of IGM at $\zi = 5.7 - 7.3$ suggest that 
the typical size of the ionized bubble is very small, $\sim 2$ comoving Mpc, and the estimated neutral hydrogen fraction is $\sim 0.6$
from the top panel of Figure 1 of \cite{2006MNRAS.365.1012F}.
Based on these results of \xhi , we conclude the neutral hydrogen fraction is relatively high, \xhi\ $= 0.3 - 0.8$
at $\zi = 7.3$ that includes the uncertainties of the various model predictions and 
the \lya\ transmission fraction estimated from the observations.

\begin{figure}
\centering
\includegraphics[width=8cm]{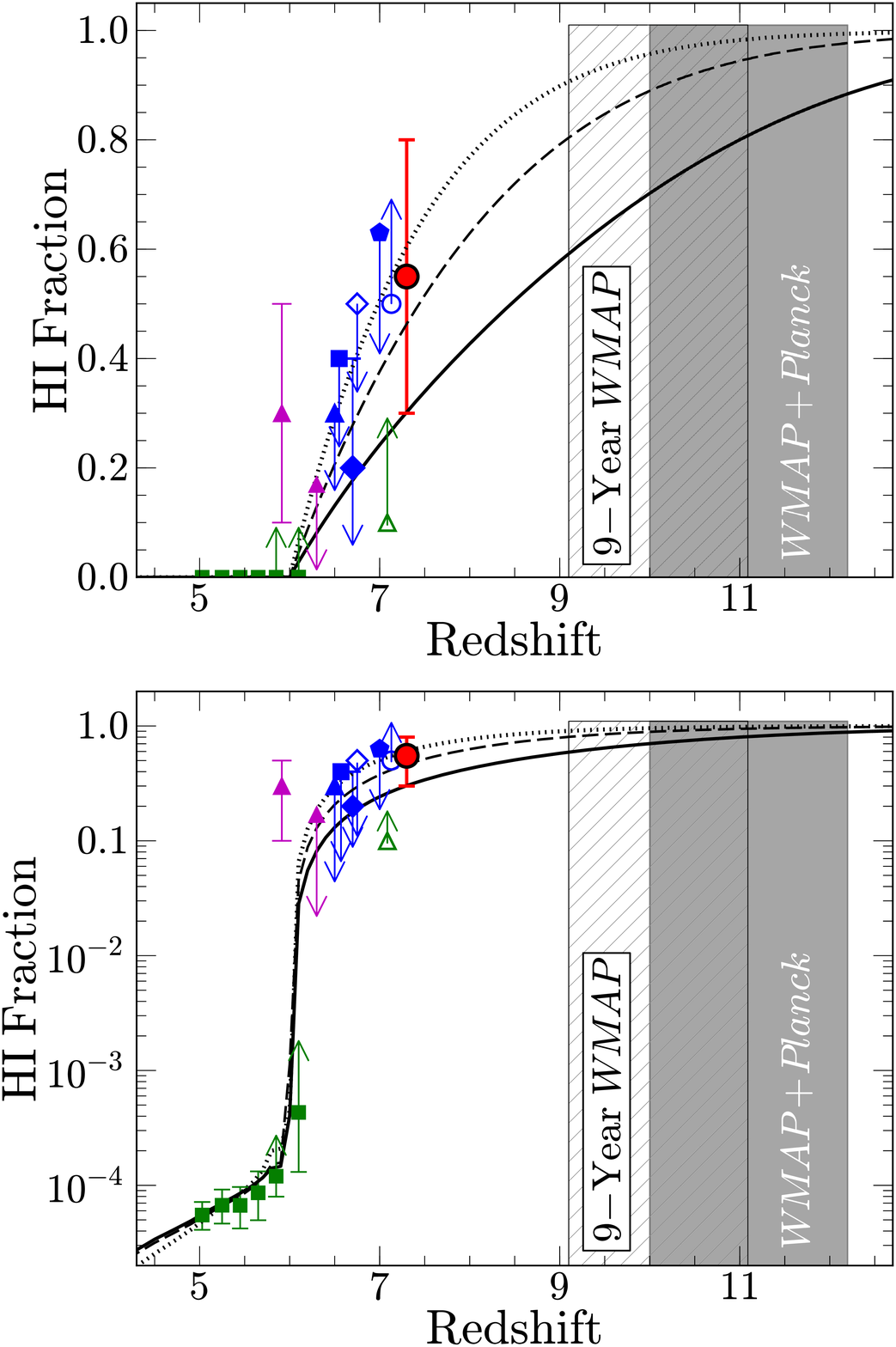}
\caption{Evolution of neutral hydrogen fraction of IGM.
Top and bottom panels are the same plots, but with the ordinate axes of linear and logarithmic scales, respectively. 
The red filled circle is the \xhi\ estimate from our \lya\ LF at $z=7.3$.
The blue filled triangle, square, diamond, and pentagon denote the \xhi\ values from the \lya\ LF evolution 
presented in \cite{2004ApJ...617L...5M}, \cite{2011ApJ...734..119K}, \cite{2010ApJ...723..869O}, 
and \cite{2010ApJ...722..803O}, respectively.
The blue open diamond and circle indicate the \xhi\ constraints given by the clustering of LAEs \citep{2010ApJ...723..869O}
and the \lya\ emitting galaxy fraction 
\citep{2011ApJ...743..132P, 2012ApJ...744..179S, 2012ApJ...744...83O, 2012ApJ...747...27T, 2012MNRAS.427.3055C, 2014MNRAS.443.2831C, 2014arXiv1403.5466P, 2014arXiv1404.4632S}, respectively.
The magenta filled triangles show the \xhi\ measurements from the optical afterglows of GRBs \citep{2006PASJ...58..485T, 2014PASJ...66...63T}.
The green filled squares and open triangle are the \xhi\ constraints provided from
the GP test of QSOs \citep{2006AJ....132..117F}  and 
the size of QSO near zone \citep{2011Natur.474..616M, 2011MNRAS.416L..70B}, respectively.
The hatched and gray regions represent the $1\sigma$ ranges for the instantaneous reionization redshifts
obtained by nine-year \textit{WMAP} \citep{2013ApJS..208...19H, 2013ApJS..208...20B}
and  \textit{WMAP}+\textit{Planck} \citep{2013arXiv1303.5076P}, respectively.
The doted, dashed and solid lines show the models A, B, and C, respectively \citep{2008MNRAS.385L..58C}.
}
\label{fig:xHI_evolution}
\end{figure}

In Figure \ref{fig:xHI_evolution}, we plot our estimate of \xhi\ at $\zi = 7.3$, and compare it with those from the previous studies.
The measurements of the \lya\ LF imply \xhi\ $< 0.63$ at $\zi = 7.0$ \citep{2010ApJ...722..803O},
and this result is consistent with our estimate of \xhi\ $ = 0.3 - 0.8$ at $\zi = 7.3$. 
The studies of \lya\ emitting fraction by \cite{2011ApJ...743..132P}, \cite{2012ApJ...744..179S}, \cite{2012ApJ...744...83O}, 
\cite{2012ApJ...747...27T}, \cite{2012MNRAS.427.3055C, 2014MNRAS.443.2831C}, \cite{2014arXiv1403.5466P}, and \cite{2014arXiv1404.4632S}
indicate \xhi\ $\gtrsim 0.5$ at $\zi \sim 7$, and 
these estimates are also comparable with ours within the uncertainties.
Moreover, the \lya\ damping wing absorption of QSO continuum
suggests \xhi\ $\gtrsim 0.1$ at $\zi = 7.1$ \citep{2011Natur.474..616M, 2011MNRAS.416L..70B}
that is, again, consistent with our estimate.

In Section \ref{sec:evolution}, we find that the decrease of the \lya\ LF at $\zi = 6.6 - 7.3$ is larger than that at $\zi = 5.7 - 6.6$.
This accelerated evolution can be also found in Figure \ref{fig:xHI_evolution}, albeit with the large uncertainties, 
by the comparison of our $\zi = 7.3$ result (red filled circle) 
with the strongest upper limit of \xhi\ from the previous $\zi = 6.6$ result (blue filled diamond).
While we find that the \lya\ LF decreases from $\zi = 6.6$ to $7.3$ at the $> 90 \%$ confidence level, 
the difference of the \xhi\ estimates between $\zi = 6.6$ and $7.3$ is only within the $1 \sigma$ level 
that is less significant than the \lya\ LF evolution result.
This is because the error bar of \xhi\ at $\zi = 7.3$ is not only from the uncertainties of 
the \lya\ LF estimates, but also from the errors of the UV LF measurements and the variance of the theoretical model
results.

It is implied that the amount of IGM neutral hydrogen may increase acceleratingly at $\zi \sim 7$.
However, the results of the \xhi\ evolution are based on various assumptions that should be
examined carefully.
In Section \ref{sec:xHI}, we assume $f^{\mathrm{esc}}_{\mathrm{Ly}\alpha, \zi = 5.7} / f^{\mathrm{esc}}_{\mathrm{Ly}\alpha, \zi = 7.3} = 1$.
Observational studies show that the \lya\ escape fraction of LAEs increases from $\zi \sim 0$ to $\sim 6$,
i.e., $f^{\mathrm{esc}}_{\mathrm{Ly}\alpha, \zi = 0} / f^{\mathrm{esc}}_{\mathrm{Ly}\alpha, \zi = 6} < 1$
(\citealt{2008ApJS..176..301O,2011ApJ...730....8H}; see also \citealt{2010ApJ...724.1524O}).
If this trend continues to $z=7.3$, the intrinsic \lya\ escape fraction with no IGM absorption
would be $f^{\mathrm{esc}}_{\mathrm{Ly}\alpha, \zi = 5.7} / f^{\mathrm{esc}}_{\mathrm{Ly}\alpha, \zi = 7.3} < 1$.
In this case, we obtain
the value of $T^{\mathrm{IGM}}_{\mathrm{Ly}\alpha, \zi = 7.3} / T^{\mathrm{IGM}}_{\mathrm{Ly}\alpha, \zi = 5.7}$
is smaller than our estimate above (see Equation \ref{eq:eq2})
and an \xhi\ estimate higher than our result of \xhi\ $= 0.3 - 0.8$ at $\zi = 7.3$.

\subsubsection{Comparison with Optical Depth of\\
Thomson Scattering} \label{sec:compare_tau}

\begin{figure}
\centering
\includegraphics[width=8cm]{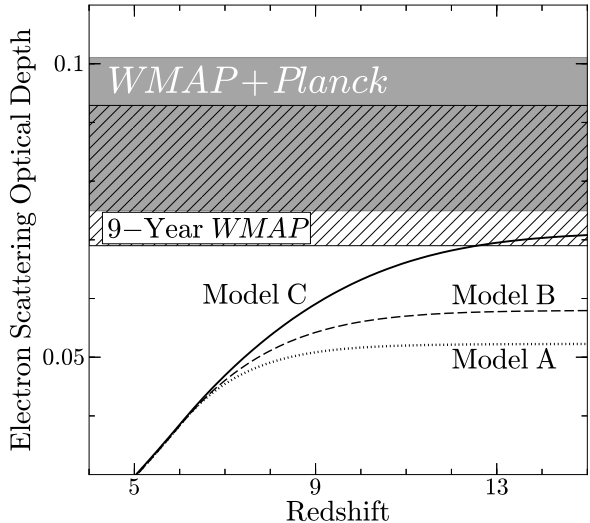}
\caption{Evolution of Thomson scattering optical depth, \tel.
The hatched and gray regions indicate the $1\sigma$ ranges of the \tel\ measurements
of \tel $=0.081 \pm 0.012$ and \tel $= 0.089^{+0.012}_{-0.014}$
obtained by nine-year \textit{WMAP} \citep{2013ApJS..208...19H, 2013ApJS..208...20B}
and \textit{WMAP}+\textit{Planck} \citep{2013arXiv1303.5076P}, respectively.
The doted, dashed, and solid curves represent the models A, B, and C, respectively \citep{2008MNRAS.385L..58C}.
} 
\label{fig:tauel}
\end{figure}

In this section, we investigate whether the relatively high value of our \xhi\ estimate 
can explain the Thomson scattering optical depth, \tel, measurements given by \textit{WMAP} and \textit{Planck}.
Because one needs to know \xhi\ evolution at $z=0-1100$ to derive \tel,
we use three models of the \xhi\ evolution \citep{2008MNRAS.385L..58C}
that cover typical scenarios of the early and relatively-late cosmic reionization history.
We refer to these three \xhi\ evolution models as models A, B, and C 
corresponding to the minimum halo masses for reionization sources that are
$\sim 10^{9}$, $\sim 10^{8}$, and $\sim 5 \times 10^{5} \ M_{\odot}$, respectively, 
at $\zi = 6$ in the semi-analytic models of \cite{2008MNRAS.385L..58C}.
We present the \xhi evolution of models A, B, and C in Figure \ref{fig:xHI_evolution},
and \tel\ as a function of redshift for these models in Figure \ref{fig:tauel}.
In Figure \ref{fig:tauel},
the hatched and gray regions represent the $1 \sigma$ range of \tel\ measured by \textit{WMAP}
and \textit{WMAP}+\textit{Planck}, respectively.
While models A and B  are consistent with our \xhi\ estimate at $z=7.3$ in Figure \ref{fig:xHI_evolution},
the models A and B fall far below 
the \tel\ measurements of \textit{WMAP} and \textit{WMAP}+\textit{Planck} in Figure \ref{fig:tauel}.
These results require reionization that proceeds at an epoch earlier than the models A and B.
The model C is such an early reionization model that just agrees with the lower end of the error of our \xhi\ estimate at $z=7.3$
in Figure \ref{fig:xHI_evolution}. However, the model C is barely consistent with the \textit{WMAP} result 
within the $1\sigma$ error in Figure \ref{fig:tauel}.
Moreover, in Figure \ref{fig:tauel}, the \tel\ value from \textit{WMAP}+\textit{Planck} is higher than the one of model C beyond the uncertainty.
Thus, there is a possible tension between our estimate of high \xhi\ and the CMB measurements of high \tel.
A similar tension between \tel\ and galaxy observation results is also claimed by \citet{2010Natur.468...49R} who discuss 
UV luminosities of reionization sources that are based on observational quantities independent from the \lya\ LFs of our study.

\subsubsection{Physical Origin of the Accelerated Evolution of \lya\ LF}
\label{sec:physical_origin}

The physical origin of the accelerated \lya\ LF evolution could be something other than the rapid increase of the neutral hydrogen at $z\gtrsim 7$,
because the \tel\ measurements have a tension with the high $x_{\rm HI}$ value that is estimated with 
our \lya\ LF under the assumption that the \lya\ LF evolution is given by the combination of 
cosmic reionization and cosmic SFR density evolution.
Similarly, large values of $x_{\rm HI}$ estimates at $z\simeq 6-7$ are obtained from 
the \lya\ damping wing absorption techniques with LAEs \citep{2006ApJ...648....7K, 
2011ApJ...734..119K,2008ApJ...677...12O,2010ApJ...722..803O,2010ApJ...723..869O, 2012ApJ...752..114S},
LBGs  \citep{2011ApJ...743..132P, 2012ApJ...744..179S, 2012ApJ...744...83O, 2012ApJ...747...27T, 2012MNRAS.427.3055C, 2014MNRAS.443.2831C, 2014arXiv1403.5466P, 2014arXiv1404.4632S},
QSOs \citep{2011MNRAS.416L..70B}, and GRBs \citep{2014PASJ...66...63T}.
Recent theoretical studies suggest a few physical pictures that explain the \tel\ measurements
and the large $x_{\rm HI}$ estimates given by the \lya\ damping wing absorption measurements.
The first picture is the presence of clumpy neutral hydrogen clouds in ionized bubbles
at the end of reionization epoch. 
\lya\ line and UV continuum from objects would be
attenuated by a number of optically thick absorption systems 
that have large {\sc Hi} column densities such as Lyman limit systems
\citep{2013MNRAS.429.1695B, 2014ApJ...781...97X}.
The absorption systems of the clumpy {\sc Hi} clouds do not contribute to
the volume-limited value of \xhi\ significantly, but to the attenuation of \lya-line and
UV-continuum emitted from objects. 
Interestingly, recent ALMA observations report a possible {\sc Hi} cloud emitting
{\sc [Cii]158}$\mu$m near a star-forming galaxy at $z=6.6$ \citep{2014arXiv1403.4360O},
supporting this physical picture.
If this picture is correct,
our finding of the accelerated \lya\ LF evolution indicates that
the number of such clumpy {\sc Hi} clouds rapidly increases 
at $z\gtrsim 7$.
The second picture is the increase of ionizing photon escape fraction towards high-$z$ \citep{2014MNRAS.440.3309D}.
\lya\ photons are produced by recombination following photoionization in ionized gas of a galaxy.
The more the ionizing photons escape from the galaxy, the smaller an amount of recombination is.
Under the significant escape of the ionizing photons, \lya\ emission is not efficiently produced by ionized clouds in
a galaxy. This picture reconciles with the increase of the ionizing photon escape fraction
suggested by \citet{2014MNRAS.442..900N} from the ionization parameter evolution.
If this picture is correct, the accelerated \lya\ LF evolution would suggest
either the sudden decrease of the gas covering fraction of galaxies or
the boosting of the ionization parameter that would make density-bounded clouds in galaxies.
However, in this picture, the high \xhi\ values given by the UV-continuum studies of QSOs and GRBs
are not explained. Additional physical mechanisms would be required for these \xhi\ results of
the UV-continuum studies.

Another possibility is that the tension between the \xhi\ estimates and
\tel\ would not exist, because the uncertainty of the \xhi\ estimates are very large (Figure \ref{fig:xHI_evolution}).
In fact, the tension is found at the significance level only beyond $\sim 1$ sigma.
It is not clear whether the tension is a hint for the discrepancy between the \xhi\ and \tel\ estimates.
One of the dominant factors of the \xhi\ uncertainty is the error of the \lya\ LFs, 
which is largely caused by the statistical errors due to the small LAE samples.
To obtain a large sample of LAEs, it is necessary to carry out narrowband imaging observations
in survey fields significantly wider than this study. 
One promising project is the Subaru/Hyper Suprime-Cam (HSC) survey that will
complete 30 deg$^2$ and 3.5 deg$^2$ narrowband observations for LAEs at $z=5.7-6.6$ and $5.7-7.3$, respectively,
with the depth comparable with those accomplished by the present Subaru surveys.
With the strong constraint of \xhi\ given by the HSC survey,
we will address the problem whether the tension is a discrepancy between \xhi\ and \tel\ estimates
and a new physical picture is really needed.

\section{Summary}\label{sec:summary}

We have conducted the ultra-deep Subaru/Suprime-Cam imaging survey for $\zi = 7.3$ LAEs with 
our custom narrowband filter, \textit{NB101}, that has a sharp bandpass for a high sensitivity 
of faint line detection.
We have observed a total of $\simeq 0.5$ deg$^2$ sky of SXDS and COSMOS fields with the integration times of 
36.3 and 69.5 hours, respectively. We have reached the $5\sigma$ limiting luminosity of 
$L(\mathrm{Ly}\alpha) \sim 2.4 \times 10^{42}$ erg $\mathrm{s}^{-1}$,
which is about 4 times deeper than those achieved by the previous Subaru studies for $\zi \gtrsim 7$ LAEs
and comparable with the luminosity limits of the previous Subaru $z=3.1-6.6$ LAE surveys.
Our observations allow us to derive the \lya\ LF at $\zi = 7.3$ with the unprecedented accuracy,
and to examine the \lya\ LF evolution from $\zi = 6.6$ to $7.3$ reliably.
The major results of our study are listed below.

\begin{enumerate}

\item
We identify three and four LAEs in SXDS and COSMOS fields, respectively.
These numbers are surprisingly small, because we expect to find
a total of $\sim 65$ LAEs by our survey in the case of no evolution of \lya\ LF from $\zi = 6.6$ to $7.3$.
We derive the \lya\ LF at $z=7.3$ with our data, carefully evaluating
uncertainties of Poisson statistics and cosmic variance.
We fit Schechter functions to our \lya\ LF, and
obtain the best-fit Schechter parameters,
$L^{*}_{\mathrm{Ly}\alpha} = 2.7^{+8.0}_{-1.2} \times 10^{42} \ \mathrm{erg} \ \mathrm{s}^{-1}$
and $\phi^{*} = 3.7^{+17.6}_{-3.3} \times 10^{-4} \ \mathrm{Mpc}^{-3}$ with a fixed $\alpha = -1.5$.

\item
We compare our \lya\ LF with the previous measurements of the \lya\ LF at $z\simeq 7.3$.
Our \lya\ LF measurements are consistent with those of the previous Subaru and VLT studies, but 
significantly smaller than those of the 4m-telescope observations. The significant differences of the \lya\ LF between the 4m-telescope
programs and Subaru+VLT studies including ours could not be explained by
cosmic variance. It is possible that the 4m-telescope results are derived with the highly contaminated
LAE samples, as suggested by the recent spectroscopic follow-up observations that
find no emission lines in the LAEs of the 4m-telescope samples.

\item
We identify the decrease of the \lya\ LF from $\zi = 6.6$ to $7.3$ significantly at the $> 90\%$ confidence level
in the Schechter function parameter space (Figure \ref{fig:error_cont}), comparing with the \lya\ LFs
at $z=6.6$ obtained from the largest LAE sample of Subaru survey
with the estimates of the cosmic variance uncertainties. 
Using our \lya\ LFs at $z=7.3$ and the Subaru results at $z=5.7-6.6$,
we find the rapid decrease of the \lya\ LF indicated by
the evolution of the \lya\ luminosity density ratios
(Figure \ref{fig:rho_evo}).
Approximating the evolution of the \lya\ luminosity density with 
the power-law function, $(1+z)^{n(\rho)}$,
we obtain $n(\rho) = -5.0^{+4.2}_{-9.5}$ at $\zi = 5.7-6.6$ and $n(\rho) = -20.8^{+5.1}_{-9.4}$ at $\zi = 6.6 - 7.3$.
Because these values of $n(\rho)$ are significantly different beyond the uncertainties,
we conclude that there is the accelerated evolution of the \lya\ LF at $z\gtrsim 7$.

\item 
Because no accelerated evolution of the UV-continuum LF or the cosmic star-formation rate (SFR)
is found at $z\sim 7$, but suggested only at $z>8$, if any \citep{2013ApJ...773...75O, 2014arXiv1403.4295B}, 
this accelerated \lya\ LF evolution is explained 
by physical mechanisms different from pure SFR decreases of galaxies
but related to the \lya\ production and escape in the process of cosmic reionization. 
We discuss the simple scenario of cosmic reionization that contributes to the accelerated evolution of the \lya\ LF. 
Subtracting the effect of the galaxies' SFR evolution from the decrease of the \lya\ luminosity density,
we estimate the ratio of \lya\ transmission of IGM
to be $T^{\mathrm{IGM}}_{\mathrm{Ly}\alpha, \zi = 7.3} / T^{\mathrm{IGM}}_{\mathrm{Ly}\alpha, \zi = 5.7} =0.29$.
By the comparison of theoretical models, we obtain \xhi\ $ = 0.3 - 0.8$ whose large uncertainty
includes the variance of the theoretical model predictions.
Although this result is consistent with previous $\zi \sim 7$ studies that use the \lya\ damping wing absorption,
there would exist a tension between the \xhi\ estimate and the Thomson scattering optical depth, \tel, measurements 
from \textit{WMAP} and \textit{Planck} at the significance level only beyond $\sim 1$ sigma.
If this tension is a hint for the discrepancies between the \xhi\ and \tel\ estimates,
these results support new physical pictures such as the clumpy neutral gas cloud absorption 
and the increase of the ionizing photon escape fraction suggested by recent theoretical studies.

\end{enumerate}

\acknowledgments

We thank Martin Haehnelt, Koki Kakiichi, Lucia Guaita, 
Masayuki Umemura, Kentaro Nagamine, and Masao Hayashi for useful comments and discussions.
We are grateful to Tirthankar Roy Choudhury for providing his data.
We appreciate Carnegie Observatories and the director,
Wendy Freedman, who provided the fund for the \textit{NB101} filter
that was key for achieving our ultra-deep survey.
This work was supported by World Premier International Research
Center Initiative (WPI Initiative), MEXT, Japan,
and KAKENHI (23244025) Grant-in-Aid for Scientific Research
(A) through Japan Society for the Promotion of Science
(JSPS). K.N. and S.Y. acknowledge the JSPS Research
Fellowship for Young Scientists.

\textit{Facility: Subaru} (Suprime-Cam)

\bibliographystyle{apj}
\bibliography{z7p3LAE_2014}

\end{document}